\documentclass[showpacs,prd,amsmath,nofootinbib,amssymb]{revtex4}
\usepackage{graphicx}
\usepackage{subfig}
\usepackage{amssymb}
\usepackage{hyperref}
\usepackage[latin1]{inputenc}
\usepackage[english]{babel}
\usepackage{epsfig}
\usepackage{wasysym}
\usepackage{color,xcolor}
\usepackage{pdflscape}
\usepackage{amsmath}
\usepackage{bm}
\usepackage{epsfig}
\usepackage{amsfonts}
\usepackage{dcolumn}

\begin{document}

\title{Bardeen-Kiselev black hole with cosmological constant\\}

\author{Manuel E. Rodrigues$^{(1,2)}$\footnote{E-mail 
address: esialg@gmail.com}, Marcos V. de S. Silva$^{(1)}$\footnote{E-mail address: marco2s303@gmail.com}, Henrique A. Vieira$^{(1)}$\footnote{E-mail 
address: henriquefisica2017@gmail.com}
}

\affiliation{$^{(1)}$Faculdade de F\'{i}sica, Programa de P\'{o}s-Gradua\c{c}\~{a}o em F\'{i}sica, Universidade Federal do Par\'{a}, 66075-110, Bel\'{e}m, Par\'{a}, Brazil\\
$^{(2)}$Faculdade de Ci\^{e}ncias Exatas e Tecnologia, Universidade Federal do Par\'{a} Campus Universit\'{a}rio de Abaetetuba, 68440-000, Abaetetuba, Par\'{a}, 
Brazil\\
}

\begin{abstract}
In this work we analyze a solution that mimics the Bardeen solution with a cosmological constant surrounded by quintessence. We show that this solution can be obtained by Einstein equations coupled with nonlinear electrodynamics. We also show that the solution is not always regular and what the conditions for regularity are. We analyze the thermodynamics associated with this type of solution by establishing the form of the Smarr formula and the first law of thermodynamics. We obtain some thermodynamic quantities such as pressure, temperature, heat capacity and isothermal compressibility. Once we have these thermodynamic quantities, we check if this solution has phase transitions and how it behaves at the points where the transitions occur. For some values of the parameters, we find that the solution exhibits a first-order phase transition, like a van der Waals fluid.
\end{abstract}

\date{\today}

\maketitle


\section{Introduction}
\label{sec1}
General relativity (GR) is able to predict and describe a number of astrophysical objects \cite{Berti:2015itd}. One of the most important predictions of GR is black holes \cite{Barack:2018yly}. Due to their causal structure, black holes have a surface to which any particle or wave passing through this surface cannot return, this surface is known as the event horizon. Recently, the study of black holes has become more important. Two events that contributed to this were the detection of gravitational waves by the LIGO/VIRGO collaboration \cite{Abbott:2016blz,Abbott:2017vtc,LIGOScientific:2018mvr,TheLIGOScientific:2016pea,TheLIGOScientific:2016wfe,TheLIGOScientific:2016htt,Abbott:2020khf}, and the first image of the environment of a black hole obtained by the Event Horizon Telescope (EHT) collaboration \cite{Akiyama:2019cqa, Akiyama:2019brx, Akiyama:2019sww,Akiyama:2019bqs,Akiyama:2019fyp,Akiyama:2019eap}. Although these experiments represent a major development in black hole research, the presence of a shadow or gravitational ringdown are not definitive proof of this type of solution \cite{Herdeiro:2021lwl,Cardoso:2016rao}.
In addition to black holes, there are a number of solutions that can be obtained through GR. Some of these solutions are, for example, wormholes \cite{Morris:1988cz,Morris:1988tu,Visser:1995cc,Lobo:2017oab,Bronnikov:2016xvj,Bronnikov:2016osp,Bronnikov:2018vbs,Blazquez-Salcedo:2020czn,Bolokhov:2021svp,Bronnikov:2021piw}, boson stars \cite{Colpi:1986ye,Liebling:2012fv,Yoshida:1997qf}, black bounce \cite{Simpson:2018tsi,Simpson:2019cer,Lobo:2020kxn,Huang:2019arj,Nascimento:2020ime,Lobo:2020ffi,Franzin:2021vnj}, and regular black holes \cite{Zaslavskii,Fan-Wang,Bronnikov:2017tnz,Capozziello:2014bqa,Rodrigues:2015,NED2,NED3,NED4,NED5,Bronnikov:2006fu,Hollenstein:2008hp,NED10,Rodrigues:2016,Rodrigues:2017,Rodrigues:2018,bambi,neves,toshmatov,DYM,ramon,berej,Rodrigues:2019,Silva:2018,Junior:2020} among others. For some of these solutions to exist, coupling with auxiliary fields such as scalar, electromagnetic, or Dirac fields are necessary. The black bounce, for example, is a type of solution that may arise from the coupling of the gravitational theory with a scalar field and an electromagnetic field \cite{Huang:2019arj}.

In the context of black holes, the problem of singularities arises, i.e., places in spacetime where the geodesics are interrupted \cite{Bronnikov:2012wsj}. Possibly, the existence of singularities is a failure of the theory because it is a classical theory of gravity. In this sense, a quantum theory of gravity could solve the problem of the existence of singularities \cite{cap2}.

We do not yet have a complete theory of quantum gravity, but there are some alternatives to eliminate the problem of the singularity. One of these alternatives is the regular black hole, so named because there are no singularities. The first regular solution was proposed by Bardeen \cite{Bardeen}. Since the Bardeen solution did not satisfy the Einstein equations for the vacuum, Beato and Garcia showed that this solution can be obtained if one takes into account the coupling of gravitational theory with nonlinear electrodynamics \cite{Beato1}. In the literature, the study of regular solutions has become quite extensive, with a large number of proposed solutions and with the analysis of the properties \cite{Macedo:2014uga,Macedo:2015qma,Macedo:2016yyo,Paula:2020yfr,Lima:2020wcb}.

Another type of solution that avoids the presence of singularities is the black bounce \cite{Simpson:2018tsi}. Unlike the regular metrics mentioned above, this type of solution has a minimal area that is not zero. Depending on the parameters of this solution, it is possible to obtain the Schwarzschild metric, a kind of regular black hole where there is a bounce instead of a singularity, and a Morris-Thorne wormhole \cite{Lobo:2020ffi}. Another novelty is that, if we consider GR, these solutions are not described only by coupling with nonlinear electrodynamics, which requires an additional matter \cite{Franzin:2021vnj,Huang:2019arj}. Bronnikov and his collaborators also proposed a solution with a structure similar to the black bounce, the black universe \cite{Bronnikov:2016xvj,Bronnikov:2016osp,Bronnikov:2006fu}. Several papers have appeared on black bounces, quasinormal modes, shadows, absorption, and others \cite{Tsukamoto:2020bjm,Zhou:2020zys,Cheng:2021hoc,Tsukamoto:2021caq,Yang:2021cvh,Bambhaniya:2021ugr,Lima:2020auu}.

Classically, black holes are objects from whose interior nothing can escape. Thus, one would expect that no radiation can emanate from them \cite{Hawking:1971vc}. By semiclassical analysis, it has been shown that black holes can indeed emit radiation and that they also have a temperature and entropy \cite{Hawking:1975vcx}. In fact, there is a whole thermodynamics associated with the study of black holes. Just as we have the standard laws of thermodynamics, there are four laws for black hole thermodynamics \cite{Bardeen:1973gs}. In particular, the first law provides us with a kind of energy conservation equation, and thus it is possible to relate the entropy of the black hole to its parameters such as mass, charge, and rotation \cite{Wald:1999vt}. Since it is a thermodynamic system, it is also possible to analyze the thermodynamic stability of black holes. For example, the Schwarzschild solution has a negative heat capacity, so it is a thermodynamically unstable solution \cite{Wald:1999vt}.

The entropy of a black hole and the area are related by $S=A/4$ \cite{Bekenstein:1973ur}. However, for some solutions, if the first law of thermodynamics is maintained, the relation between area and entropy is no longer preserved \cite{Kumar:2020cve}. Since entropy can be determined by Wald's formula \cite{Wald:1999vt} and we consider general relativity, the relation between entropy and the area of the black hole is not changed. In this way, the first law of thermodynamics must be modified depending on the type of solution we consider \cite{Ma:2014qma,Zhang:2016ilt}. This type of modification can affect the thermodynamic stability of a solution, since its temperature and heat capacity also change \cite{Maluf:2018lyu}.

The structure of this paper is as follows. In Sec. \ref{sec2} we introduced and analyze some general properties, such as the spacetime conditions of regularity. In Sec. \ref{sec3} we start analyzing the thermodynamics of this solution using the temperature, the Smarr formula, and the first law of thermodynamics. Section \ref{sec4} is devoted to the study of the heat capacity and its importance for thermodynamic stability. In Sec. \ref{sec5} we obtain the equation of state and the possibility to analyze this solution from the point of view of virtual micromolecules. In Sec. \ref{sec6} we analyze the Helmholtz free energy and the isothermal compressibility to compare them with the heat capacity results. In Sec. \ref{sec7} we see how the thermodynamic functions behave near the critical points to obtain the critical exponents. Our conclusions and perspectives can be found in Sec. \ref{sec8}.

In this paper we consider natural units, where $c=\hbar=G=1$, and the metric signature $(+,-,-,-)$. We adopt the convention that Greek indices run from $0$ to $3$, so that $x^0=t$, $x^1=r$, $x^2=\theta$, and $x^3=\phi$.


\section{Bardeen-Kiselev solution with cosmological constant}
\label{sec2}
We consider a spherically symmetric spacetime
\begin{equation}
ds^2=f(r)dt^2-\frac{1}{f(r)}dr^2-r^2\left(d\theta^2+\sin^2\theta d\phi^2\right).
\end{equation}

The solution is magnetically charged and has cosmological constant. The action that describes this theory is
\begin{equation}
S=\int d^4x\sqrt{-g}\left[R+2\lambda+L_{ NED }\right],
\end{equation}
where $R$ is the curvature scalar, $\lambda$ is the cosmological constant, and $L_{ NED }$ is the nonlinear Lagrangian of electromagnetic theory. The Lagrangian is a nonlinear function of the electromagnetic scalar $\mathcal{F}=F^{\mu\nu}F_{\mu\nu}$, where $F^{\mu\nu}$ is the Maxwell-Faraday tensor.

For a spherically symmetric spacetime that is only magnetically charged, the only nonzero component of $F_{\mu \nu}$ is \cite{NED4} \footnote{In \cite{PlebanskiKrasinski}, the authors explicitly show how the nonzero components are obtained for a spherically symmetric electromagnetic field.}
\begin{equation}
F_{23}=q \sin\theta,
\end{equation} 
and the scalar $\mathcal{F}$ is
\begin{equation}
\mathcal{F}=\frac{2q}{r^4},
\end{equation}
where $q$ is the magnetic charge.

Let us consider the Lagrangian
\begin{equation}
L(\mathcal{F})=\frac{24 \sqrt{2} M q^2}{\kappa ^2 \left( \sqrt{\frac{2 q^2}{\mathcal{F}}}+2 q^2\right)^{5/2}}-\frac{6\omega c \left(\frac{2\mathcal{F}}{q^2}\right)^{\frac{3}{4} (\omega +1)}}{\kappa ^2}.
\end{equation}
When $c\rightarrow 0$, the Bardeen solution is recovered. In the limit $\mathcal{F}\rightarrow 0$ 

\begin{equation}
L(\mathcal{F})\approx -\frac{6 c 2^{\frac{3}{4} (\omega +1)} \omega \mathcal{F}^{\frac{3 (\omega +1)}{4}} }{\kappa ^2\left(q^2\right)^{\frac{3}{4} (\omega +1)}}+\frac{12 \sqrt [4]{2}\mathcal{F} \mathcal{F}^{1/4} m}{\kappa ^2 \sqrt [4]{q^2}}+O\left(\mathcal{F}\mathcal{F}^{3/4}\right).
\end{equation}
When $\omega=1/3$, we find
\begin{equation} L(\mathcal{F})\approx -\frac{4 c \mathcal{F} }{\kappa ^2 q^2}+\frac{12 \sqrt [4]{2}\mathcal{F} \mathcal{F}^{1/4} m}{\kappa ^2 \sqrt [4]{q^2}}+O\left(\mathcal{F}\mathcal{F}^{3/4}\right),
\end{equation}
so that for very small values of $\mathcal{F}$ the linear term dominates,
\begin{equation}
L(\mathcal{F})\approx \mathcal{F}, \mathcal{F}\rightarrow 0.
\end{equation}
This means that electromagnetism in the Maxwell limit behaves approximately like Maxwell theory.

In the presence of the cosmological constant the Einstein equations are
\begin{equation}
R_{\mu\nu}-\frac{1}{2}g_{\mu\nu}R+\lambda g_{\mu\nu}=\kappa^2 T_{\mu\nu}.
\end{equation}
Using the stress-energy tensor to nonlinear electrodynamics
\begin{equation}
T_{\mu\nu}=g_{\mu\nu}L(\mathcal{F})-\frac{dL}{d\mathcal{F}}F_{\mu}^{\ \alpha}F_{\nu\alpha},
\end{equation}
and solving the Einstein equations, we obtain the metric function
\begin{equation}
f(r)=1-2 c r^{-3 \omega -1}-\frac{2 M r^2}{\left(q^2+r^2\right)^{3/2}}-\frac{\lambda r^2}{3}.\label{fBK}
\end{equation}
For the limit $c\rightarrow 0$ we find the Bardeen-(anti-)de Sitter solution and for $q\rightarrow0$ we find the Kiselev-(anti-)de Sitter solution with cosmological constant \cite{Malakolkalami:2015cza}.

\begin{figure}
	\includegraphics[scale=.6]{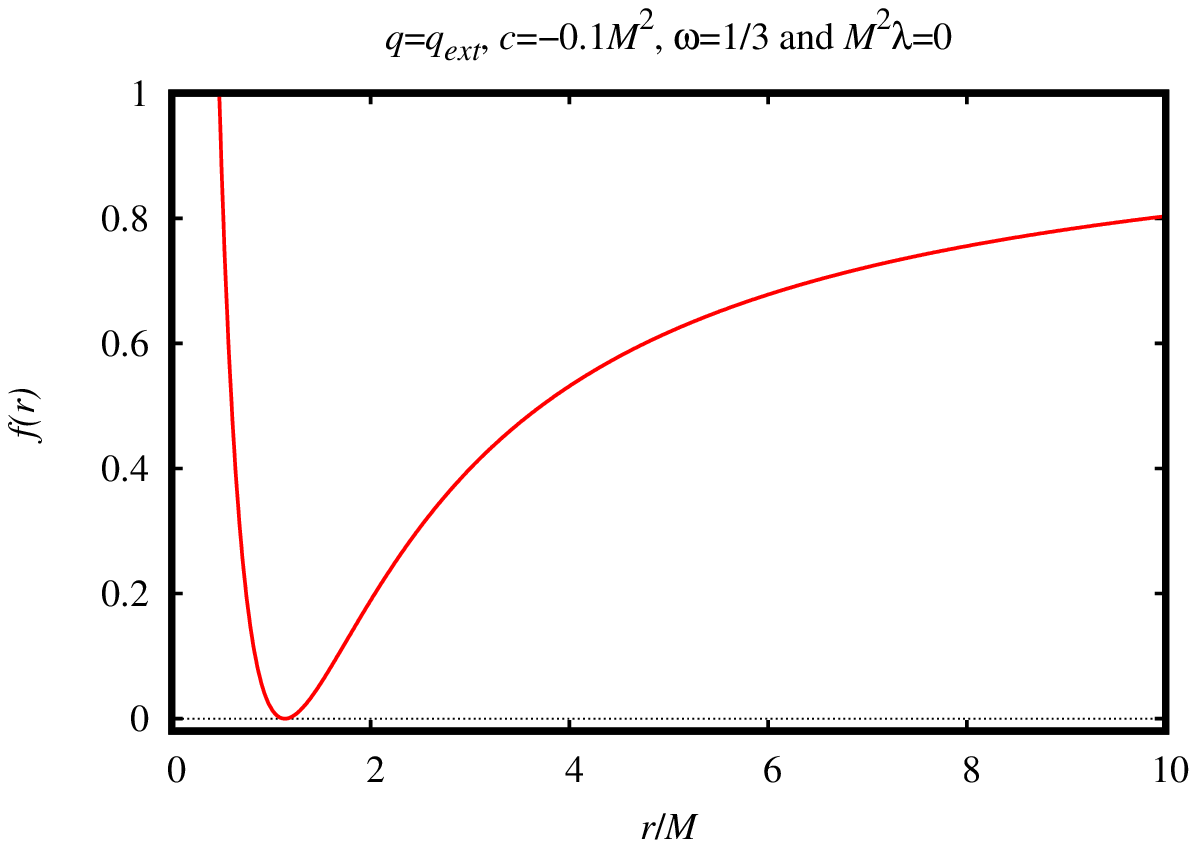}
	\includegraphics[scale=.6]{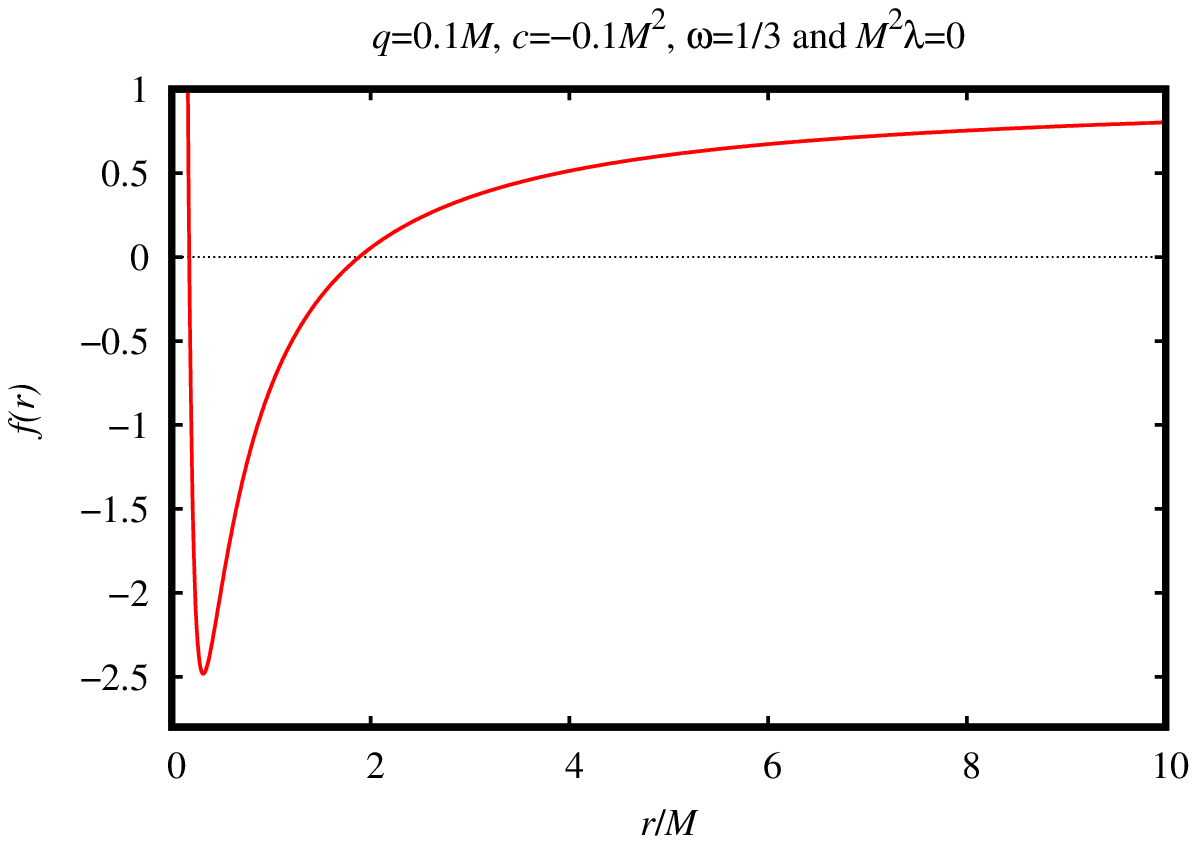}
	\includegraphics[scale=.6]{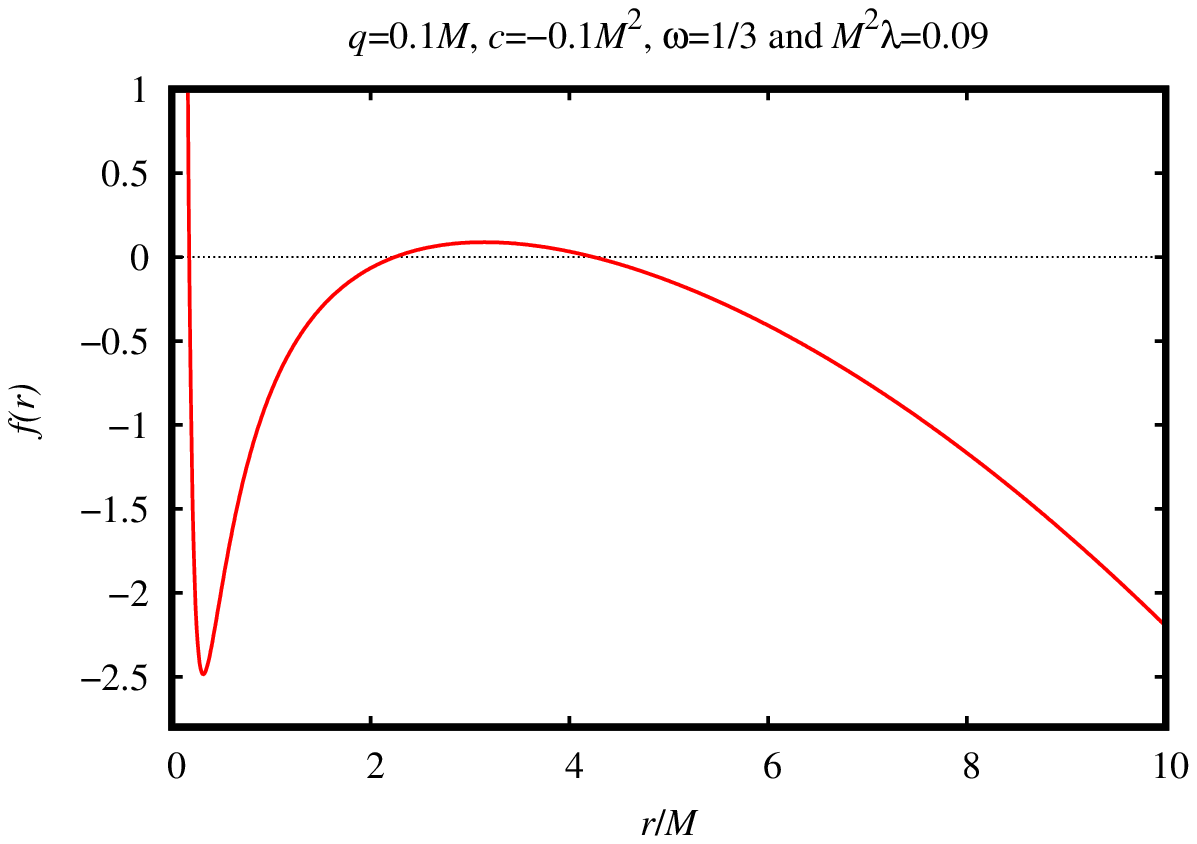}
	\includegraphics[scale=.6]{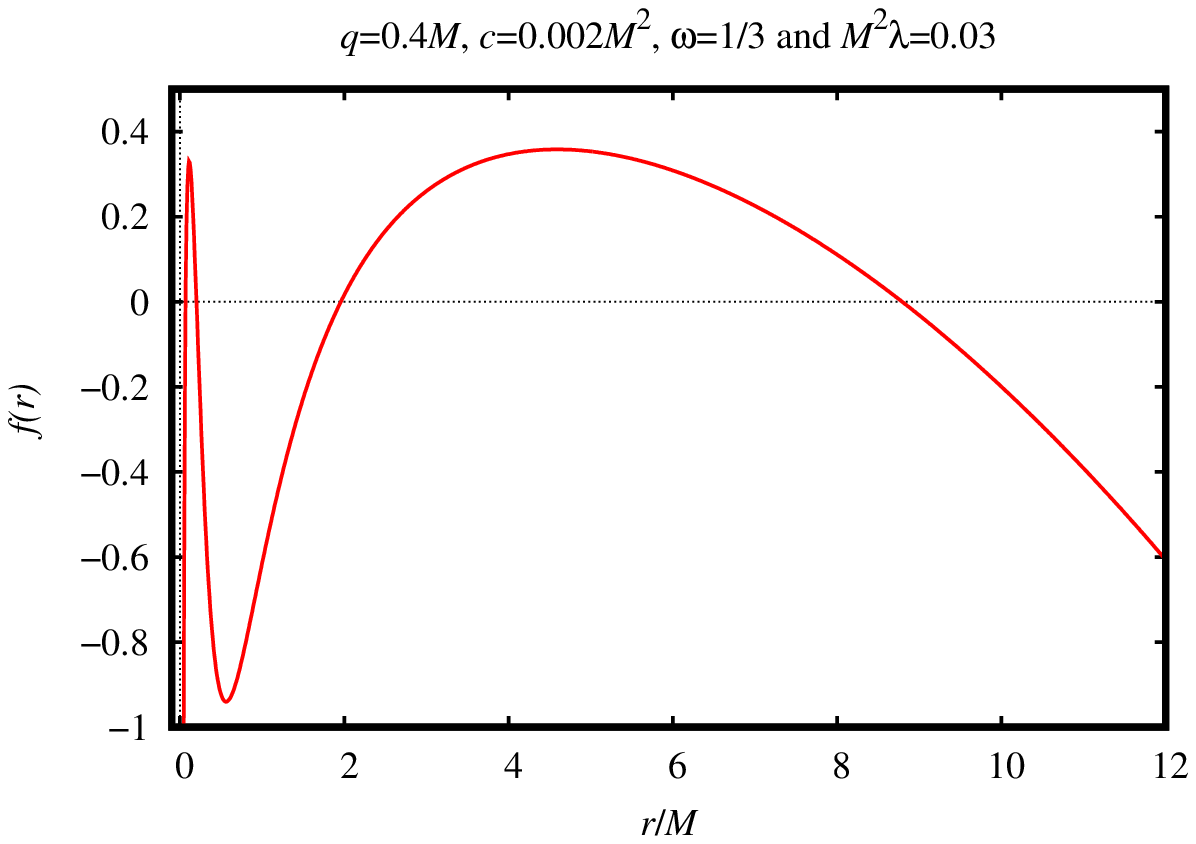}
	\caption{graphic representation of $f(r)$ with respect to the radial coordinate with different values of $q$, $\lambda$, $c$, and $\omega$.}
	\label{fig1}
\end{figure}

To obtain the horizons, we need to solve $f(r)=0$. However, we cannot obtain analytical expressions for the radius of the horizon. The number of horizons depends on the values of the parameters. In Fig. \ref{fig1} we see that the number of horizons can be up to four.

If we want to check the existence of curvature singularities, since spacetime is spherically symmetric, we only need to check the Kretschmann scalar \cite{Lobo:2020ffi}, $K=R^{\mu\nu\alpha\beta}R_{\mu\nu\alpha\beta}$, which is given by

\begin{eqnarray}
K&=&\frac{4 \left(2 c r^{-3 \omega -1}+\frac{1}{3} r^2 \left(\lambda +\frac{6
		M}{\left(q^2+r^2\right)^{3/2}}\right)\right)^2}{r^4}+\frac{4 \left(2 c (3 \omega +1) r^{-3 \omega -2}+\frac{2 M
		r \left(r^2-2 q^2\right)}{\left(q^2+r^2\right)^{5/2}}-\frac{2 \lambda r}{3}\right)^2}{r^2}\nonumber\\
	&+&\left(2 c (9 \omega 
(\omega +1)+2) r^{-3 (\omega +1)}+\frac{2}{3} \left(\lambda +\frac{M \left(-33 q^2 r^2+6 q^4+6
	r^4\right)}{\left(q^2+r^2\right)^{7/2}}\right)\right)^2.
\end{eqnarray}
The spacetime is regular only for some values of $\omega$. When $\omega\leq -1$, the solution is regular in $r\rightarrow 0$ and to $\omega\geq-1$ it is regular in $r\rightarrow \infty$ but is not asymptotically flat. For $\omega=-1$ we have the following limits:
\begin{eqnarray}
&&\lim\limits_{r\rightarrow \infty}K(r)=\frac{8}{3} (6 c+\lambda )^2,\\
&&\lim\limits_{r\rightarrow 0}K(r)=\frac{8 \left(12 M \left(q^2\right)^{3/2} (6 c+\lambda )+q^6 (6 c+\lambda )^2+36 M^2\right)}{3 q^6}.
\end{eqnarray}
For this value of $\omega$, the solution is curvature regular everywhere, since we have only a shift in the cosmological constant.

\section{Thermodynamics}\label{sec3}
To start the study of black hole thermodynamics, we can obtain the temperature of a black hole using the surface gravity by \cite{Wald:1999vt}
\begin{equation}
T_k=\frac{k}{2\pi},
\end{equation}
where
\begin{equation}
k=\left.\frac{f'(r)}{2}\right|_{r=r_+},
\end{equation}
is the surface gravity and $r_+$ is the event horizon radius. Considering Eq. \eqref{fBK}, we obtain
\begin{equation}
T_k=\frac{1}{4 \pi}\left(2 c (3 \omega +1) r_+^{-3 \omega -2}+\frac{2 M r_+ \left(r_+^2-2 q^2\right)}{\left(q^2+r_+^2\right)^{5/2}}-\frac{2
		\lambda r_+}{3}\right).\label{Tk}
\end{equation}

The first law of black hole thermodynamics for a charged and static black hole is given by \cite{Gunasekaran:2012dq}
\begin{equation}
dM=T_HdS + \Phi dq+VdP,\label{flW}
\end{equation}
where $M$ is the total energy of the system, $\Phi$ is the magnetic potential, $P$ is the thermodynamic pressure, written as
\begin{equation}
P=-\frac{\lambda}{8\pi},
\end{equation}
and $S$ is the entropy of the system. The area law states that \cite{Bekenstein:1973ur}
\begin{equation}
S=\frac{A}{4}=\pi r_+^2,\label{Entropy}
\end{equation}
where $A$ is the area of the black hole.

The first law, as given in Eq. \eqref{flW}, is not valid if we have nonlinear electrodynamics \cite{Ma:2014qma,Zhang:2016ilt,Maluf:2018lyu}. The temperature and magnetic potential can be derived from the first law as follows:
\begin{equation}
T_H=\frac{\partial M}{\partial S},\ \mbox{and} \ \Phi=\frac{\partial M}{\partial q}.\label{Temperaturewrong}
\end{equation}
We can obtain $M$ from the condition $f(r_+)=0$, which leads to
\begin{equation}
M=-\frac{1}{6} \left(q^2+r_{+}^2\right)^{3/2} r_{+}^{-3 \omega -3} \left(6 c+\lambda r_{+}^{3 \omega +3}-3 r_{+}^{3 \omega+1}\right).\label{mass}
\end{equation}
The temperature in \eqref{Temperaturewrong}, with \eqref{mass}, is given by
\begin{equation}
T_H=\frac{\sqrt{q^2+r_{+}^2} r_{+}^{-3 \omega -5} \left(6 c \left(q^2 (\omega +1)+r_{+}^2 \omega \right)+r_{+}^{3 \omega +1} \left(-2
	q^2-\lambda r_{+}^4+r_{+}^2\right)\right)}{4 \pi }.
\end{equation} 
The temperature $T_k$ is different from $T_H$. In addition to temperature, the others quantities derived from the usual first law of thermodynamics, derivative of the mass with respect to some parameter, also present problems. To solve this problem, we need to modify the first law. The usual first law arises in a context where the Lagrangian of the theory does not explicitly depend on the mass of the black hole. Thus, when building the first law for solutions with nonlinear electrodynamics, derivatives of the stress-energy tensor must be taken into account and these corrections must modify the first law in such a way that the old thermodynamic quantities must relate to the new ones through of a correction factor.

The new first law is written as \cite{Singh:2020xju}
\begin{equation}
d\mathcal{M}=TdS + \Phi dq+PdV,\label{fltr}
\end{equation}
where
\begin{equation}
d\mathcal{M}=dM\left(1+4\pi \int_{r_+}^{\infty}r^2\frac{\partial T^0_{\ 0}}{\partial M}dr\right)=W(r_+,q)dM,
\end{equation}
where $W(r_+,q)$ is the correction factor, given by
\begin{equation}
W(r_+,q)=\left(1+4\pi \int_{r_+}^{\infty}r^2\frac{\partial T^0_{\ 0}}{\partial M}dr\right),
\end{equation}
and $T^{0}_{\ 0}$ is one of the components of the stress-energy tensor.

We find that the relation between the temperatures is
\begin{equation}
T_k=W\left(r_+,q\right)T_H=W\left(r_+,q\right)\frac{\partial M}{\partial S}=\frac{r_{+}^{-3 \omega -2} \left(6 c \left(q^2 (\omega +1)+r_{+}^2 \omega \right)+r_{+}^{3 \omega +1} \left(-2 q^2-\lambda 
	r_{+}^4+r_{+}^2\right)\right)}{4 \pi \left(q^2+r_{+}^2\right)},
\end{equation}
and $W(r_+,q)$ is 
\begin{equation} W(r_+,q)=\frac{r_{+}^3}{\left(q^2+r_{+}^2\right)^{3/2}}.\label{correctionfactor}
\end{equation}

When we have an extreme black hole, as we see in the first image of Fig. \ref{fig1}, the temperature arising from surface gravity is zero. To understand this we just need to realize that the radius of the horizon is just where we have a minimum of the function $f(r)$. So the derivative at that point is zero. The correction factor is not zero for an extreme black hole. So, both $T_k$ and $T_H$ are zero for extreme black holes.

The temperatures for Bardeen-(anti-)de Sitter and Kiselev-(anti-)de Sitter are given by
\begin{eqnarray}
T_{kB}&=&-\frac{2 q^2+\lambda r_+^4-r_+^2}{4 \pi q^2 r_++4 \pi r_+^3},\\
T_{kK}&=&\frac{6 c \omega r_+^{-3 \omega -1}-\lambda r_+^2+1}{4 \pi r_+ }.
\end{eqnarray}

\begin{figure} \centering \includegraphics[scale=0.6]{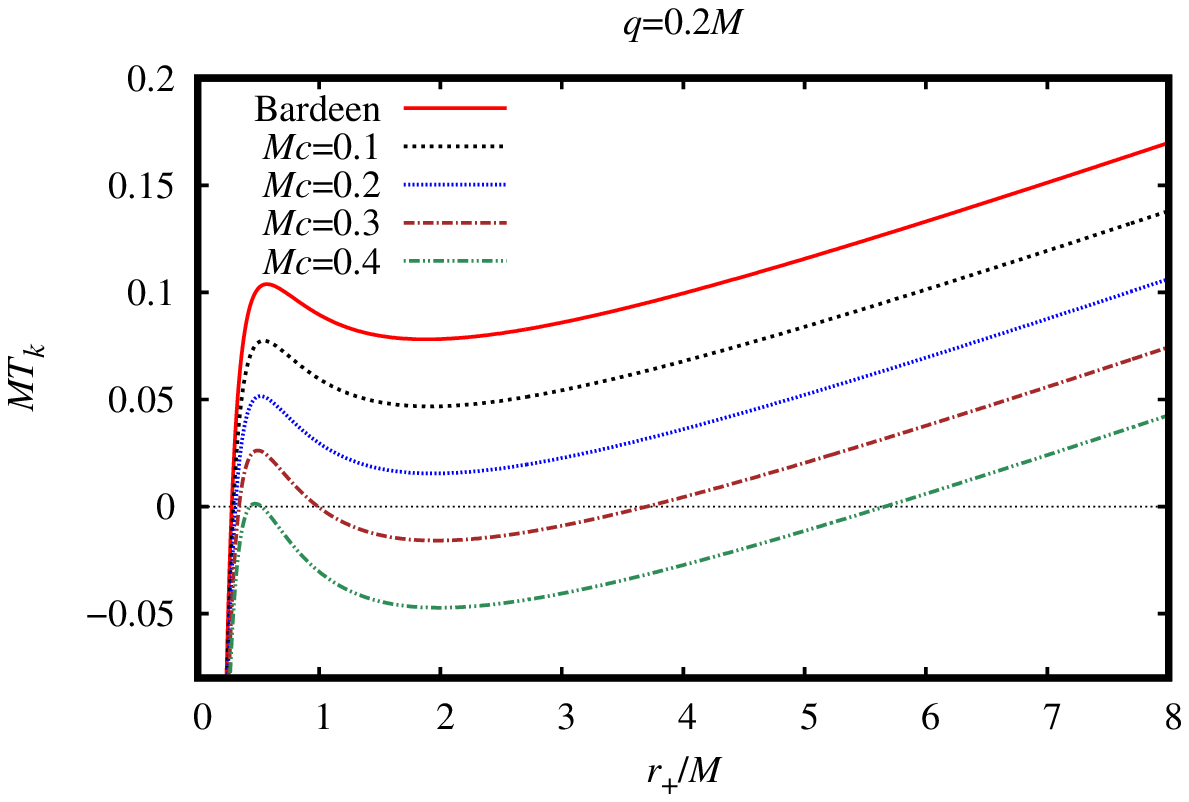} \includegraphics[scale=0.6]{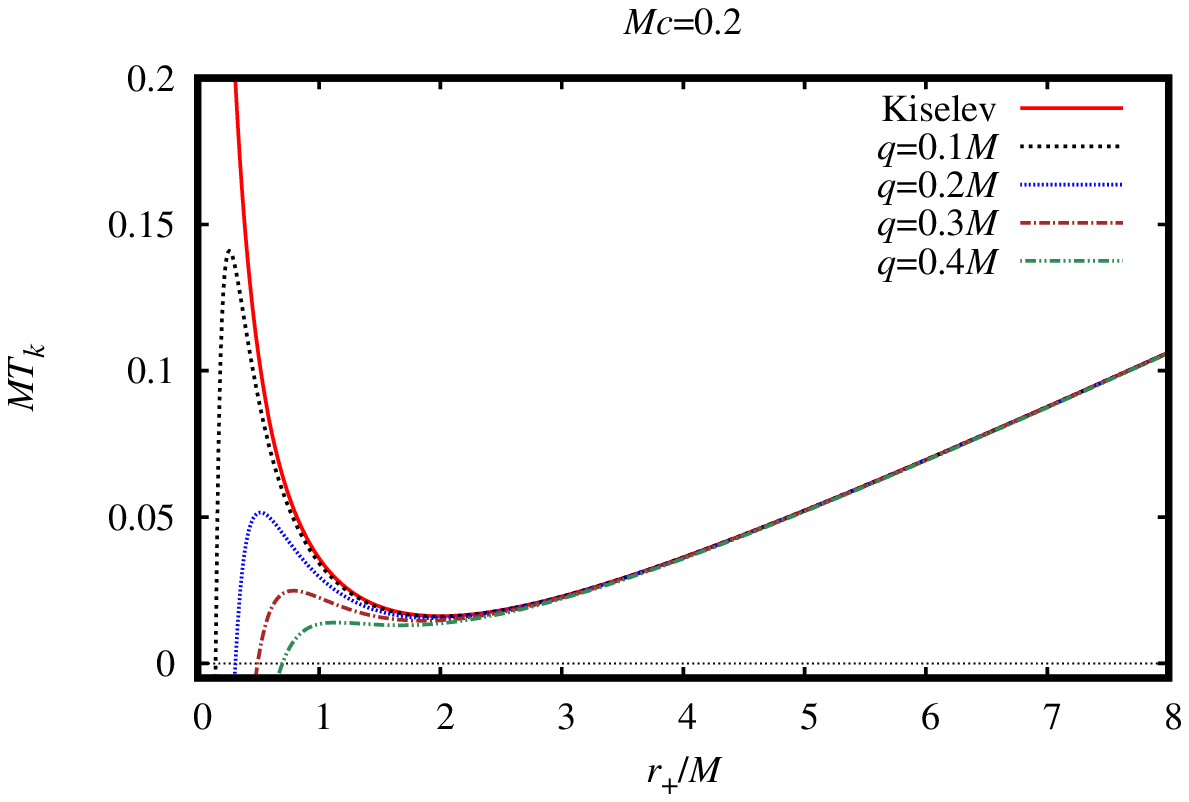} \caption{Graphic representation of the temperature associated to the solution \eqref{fBK} with $M^2\lambda=-0.08\pi$ and $\omega=-2/3$.} \label{FigTemperature}
\end{figure}
It is not so obvious what the differences are between this temperature and the temperature of Eq. \eqref{Tk}. To better analyze this situation, let us see how the temperature behaves when we change certain parameters in Fig \ref{FigTemperature}. If we assign a value to $r_+$, the temperature decreases as the charge and the parameter $c$ increase. These results show that the solution\eqref{fBK} is colder than the Bardeen- (anti-)de Sitter solution and the Kiselev-(anti-)de Sitter solution.

To obtain the complete first law, we must first find the Smarr formula, a relation that connects the mass (energy) to the other parameters through a homogeneous function. We can derive the Smarr formula from the properties of a homogeneous function \cite{Homogenea}.

In terms of the entropy, mass can be written as
\begin{equation}
M(S, q, c, \lambda)=\frac{1}{6} \left(q^2+\frac{S}{\pi }\right)^{3/2} S^{-\frac{3}{2} (\omega +1)} \left((3 \pi -\lambda S) S^{\frac{3 \omega }{2}+\frac{1}{2}}-6 c \pi ^{\frac{3 (\omega +1)}{2}}\right).\label{massS}
\end{equation}
To determine the degree of homogeneity, we write
\begin{equation}
M(l^aS, l^bq, l^dc, l^h\lambda)=\frac{1}{6} \left(l ^{2b}q^2+\frac{l ^a S}{\pi }\right)^{3/2} l ^{-\frac{3a}{2} (\omega +1)} S^{-\frac{3}{2} (\omega +1)} \left(\left(3 \pi -l^{h}\lambda l^a S\right) l^{\frac{a}{2} (3\omega +1)}S^{\frac{1}{2} (3\omega +1)}-6 \pi ^{\frac{3 (\omega +1)}{2}} l^{d}c\right).
\end{equation}
To isolate $l$, we assume $h=-a$, $d=a(3 \omega +1)/2$, $b=a/2$, and $a=1$, which leads to
\begin{equation}
M(lS, l^{1/2}q, l^{(3 \omega +1)/2}c, l^{-1}\lambda)=-\frac{1}{6} \sqrt{l} \left(q^2+\frac{S}{\pi }\right)^{3/2} S^{-\frac{3}{2} (\omega +1)} \left(6 \pi^{\frac{3 (\omega +1)}{2}} c+\left(\lambda S-3 \pi \right) S^{\frac{1}{2} (3 \omega +1)}\right),\label{Masshomogeneous}
\end{equation}
such that the mass is a homogeneous function with degree of homogeneity $n=1/2$ \cite{Homogenea}.

Euler's identity states that for a homogeneous function with degree $n$, we have \cite{Homogenea}
\begin{equation}
n.f(x_1,x_2 ... x_{m})= a_{1}x_{1}\frac{\partial f}{\partial x_{1}}+a_{2}x_{2}\frac{\partial f}{\partial x_{2}}+a_{3}x_{3}\frac{\partial f}{\partial x_{3}}+...+a_{m}x_{m}\frac{\partial f}{\partial x_{m}}.
\end{equation}
Thus we obtain
\begin{equation}
\frac{1}{2}M(S, q, c, \lambda)= T_H S+\frac{1}{2}\Phi q+\frac{1}{2}\left(1+3\omega\right)A_c c-A_{\lambda}\lambda,\label{Smarrcorrect}
\end{equation}
where $T_H$ and $\Phi$ are given by \eqref{Temperaturewrong}, and $A_c$ and $A_{\lambda}$ are
\begin{equation}
A_c=\frac{\partial M}{\partial c}\ \mbox{and} \ A_{\lambda}=\frac{\partial M}{\partial \lambda}.\label{Smarrrelations}
\end{equation}
With Eq. \eqref{Temperaturewrong} and \eqref{Smarrrelations} we get that \eqref{Smarrcorrect} and \eqref{massS} are equal.

The first law is written as
\begin{equation}
d\mathcal{M}=W(S,q) dM=T_k dS+ W(S,q)\Phi dq+W(S,q)A_cdc+W(S,q)A_{\lambda}d\lambda.\label{FistLaw}
\end{equation}

For regular black holes, it is normal that the Lagrangian depends explicitly on some parameters, such as charge and mass. Fluctuations in the matter sector are not a problem because they are compensated by fluctuations in the geometry sector by the first law, \eqref{FistLaw}.


\section{Heat Capacity}\label{sec4}
From thermodynamics, we know that the stability of a thermodynamic system requires $C_P\geq C_V\geq0$ and $k_T\geq k_S\geq0$, where $C_P$ and $C_V$ are the heat capacity at constant pressure and volume, respectively, and $k_T$ and $k_S$ are the isothermal compressibility and isentropic compressibility, respectively \cite{Stanley}.

In black thermodynamics, compressibility and heat capacity also give us information about the stability of the system. The isothermal compressibility and heat capacity at constant pressure are given by \cite{Davies:1977bgr}
\begin{eqnarray}
C_P=\left.T_k\frac{\partial S}{\partial T_k}\right|_P,\label{capacity}\\
k_T=-\left.\frac{1}{V}\frac{\partial V}{\partial P}\right|_T.\label{compress}
\end{eqnarray}

We will focus on the heat capacity. We get that the heat capacity at constant pressure is
\begin{eqnarray}
C_P&=&\left\{2 S \left(\pi  q^2+S\right) \left(S^{\frac{3 \omega }{2}+\frac{1}{2}} \left(2 \pi ^2 q^2+\lambda  S^2-\pi  S\right)-6 c \pi ^{\frac{3 (\omega +1)}{2}} \left(\pi  q^2 (\omega +1)+S
\omega \right)\right)\right\}\nonumber\\
&\times&\Big\{6 c \pi ^{\frac{3 (\omega +1)}{2}} \left(\pi  q^2 S (\omega  (6 \omega +7)+4)+\pi ^2 q^4 (\omega +1) (3 \omega +2)+S^2 \omega  (3 \omega +2)\right)\nonumber\\
&+&S^{\frac{3	\omega }{2}+\frac{1}{2}} \left(\pi  S^2 \left(3 \lambda  q^2+1\right)-7 \pi ^2 q^2 S-2 \pi ^3 q^4+\lambda  S^3\right)\Big\}^{-1}\label{heat}
\end{eqnarray}
This result is valid only if we derive temperature from the surface gravity. If we use \eqref{Temperaturewrong}, we get
\begin{equation}
\bar{C}_P=\frac{2 S \left(\pi q^2+S\right) \left(S^{\frac{3 \omega }{2}+\frac{1}{2}} \left(2 \pi ^2 q^2+\lambda S^2-\pi S\right)-6 c \pi ^{\frac{3 (\omega +1)}{2}} \left(\pi q^2 (\omega +1)+S
	\omega \right)\right)}{6 c \pi ^{\frac{3 (\omega +1)}{2}} \left(2 \pi q^2 S \left(3 \omega ^2+5 \omega +2\right)+\pi ^2 q^4 \left(3 \omega ^2+8 \omega\hspace{-0.1cm} +\hspace{-0.1cm}5\right)+S^2 \omega (3 \omega
	+2)\right)+S^{\frac{3 \omega }{2}+\frac{1}{2}} \left(\pi S^2\hspace{-0.1cm}-\hspace{-0.1cm}4 \pi ^2 q^2 S\hspace{-0.1cm}-\hspace{-0.1cm}8 \pi ^3 q^4\hspace{-0.1cm}+\hspace{-0.1cm}\lambda S^3\right)}.
\end{equation}
These two heat capacities are connected by
\begin{equation}
C_P=W(S,q)\frac{\partial T_H}{\partial T_k}\bar{C}_P.\label{CCrelation}
\end{equation}

\begin{figure}
	\includegraphics[scale=.6]{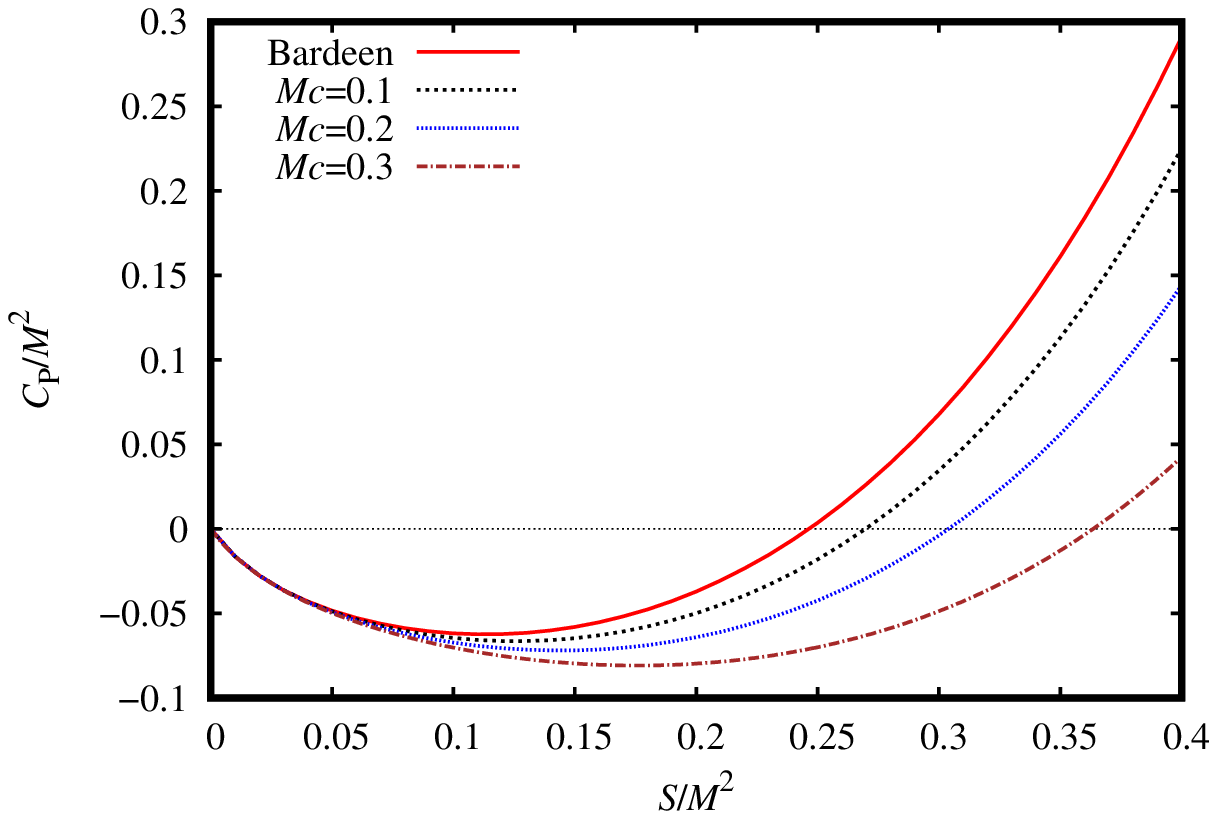}
	\includegraphics[scale=.6]{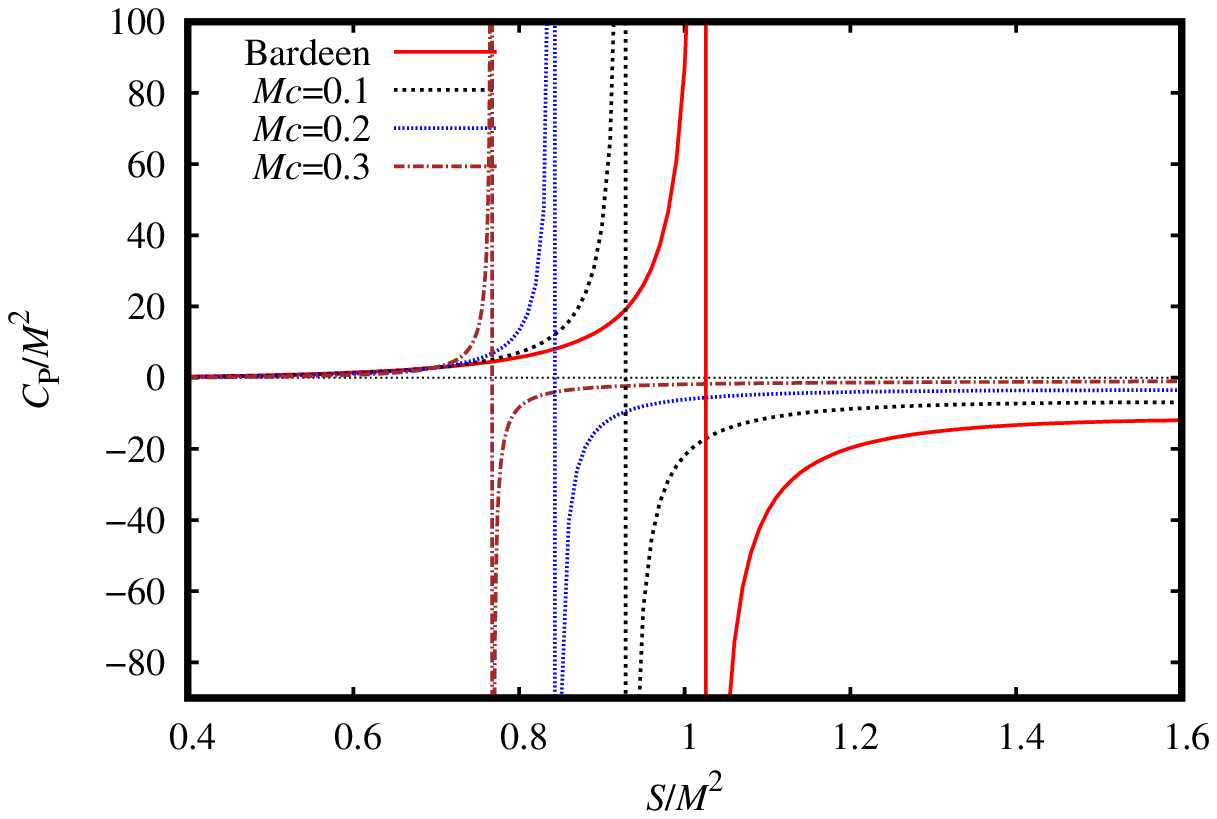}
	\includegraphics[scale=.6]{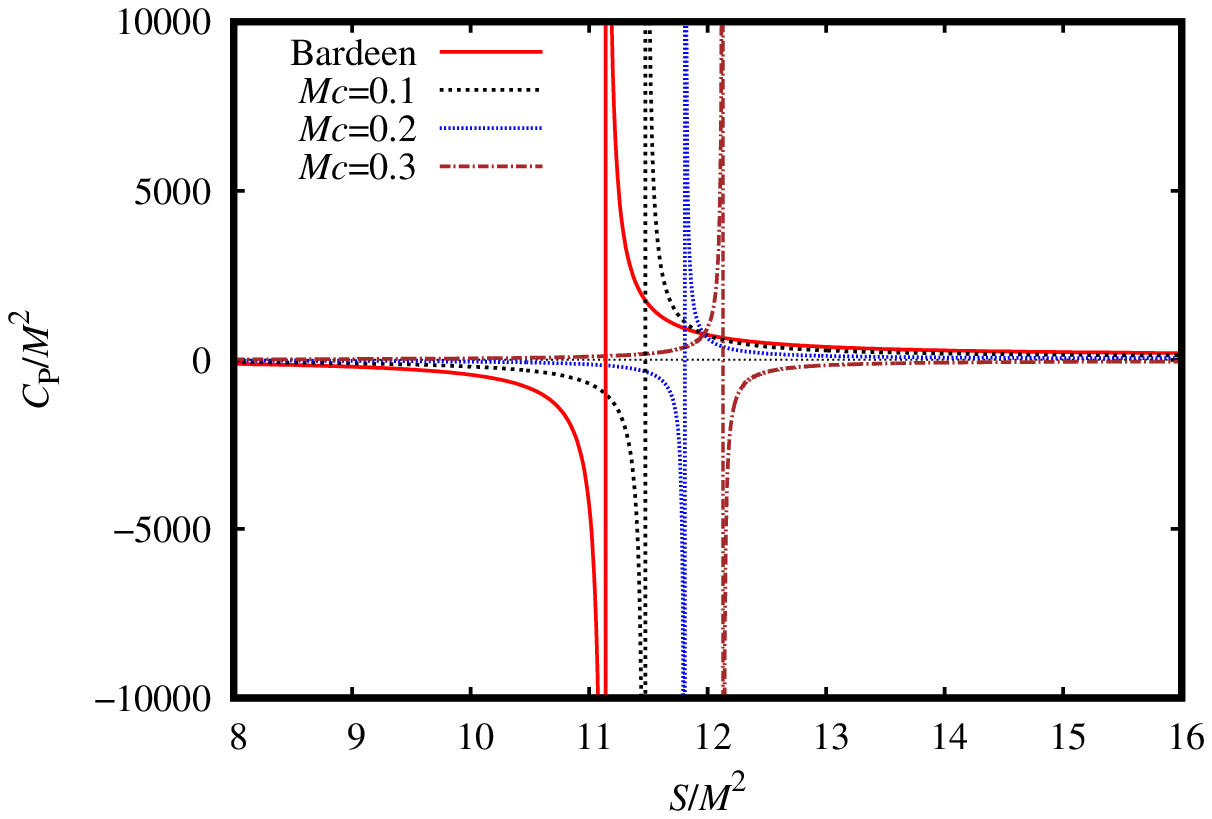}
	\includegraphics[scale=0.6]{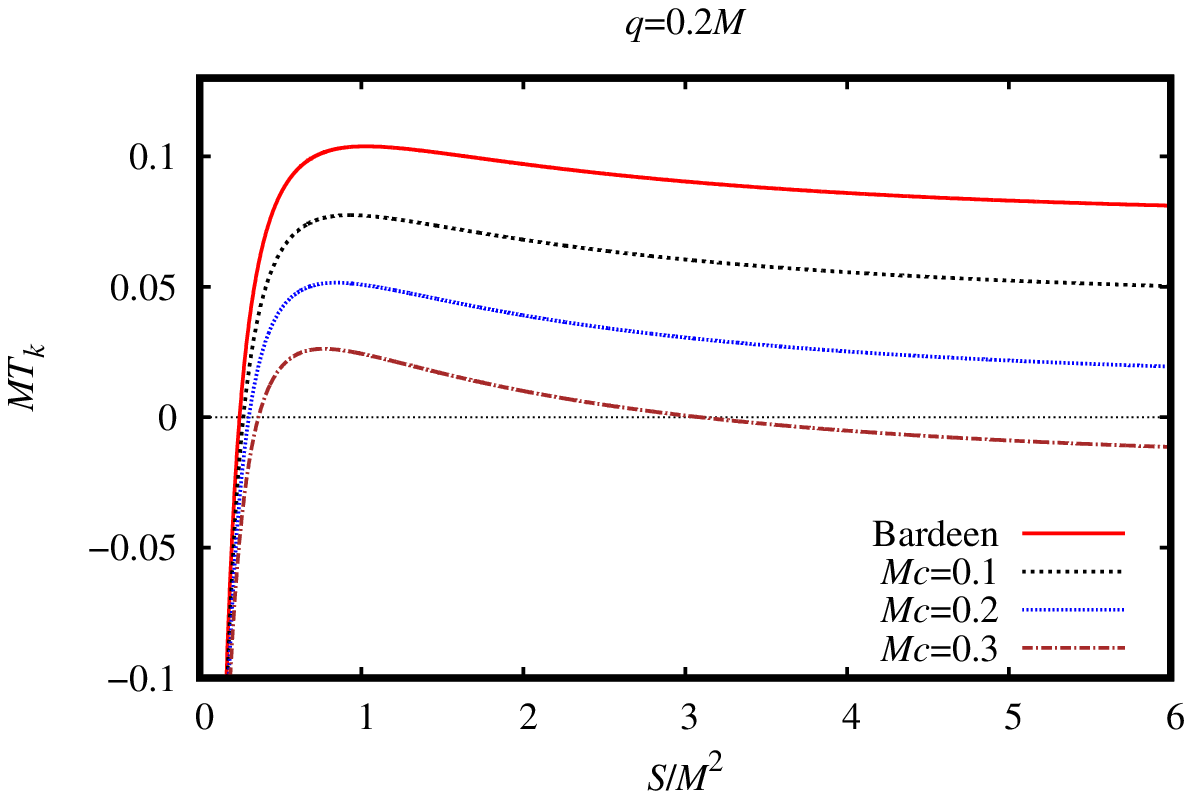}
	\caption{Heat capacity at constant pressure and temperature associated to the solution \eqref{fBK} with $q=0.2M$, $M^2\lambda=-0.08\pi$ and $\omega=-2/3$.}
	\label{figC1}
\end{figure}

\begin{figure}
	\includegraphics[scale=.6]{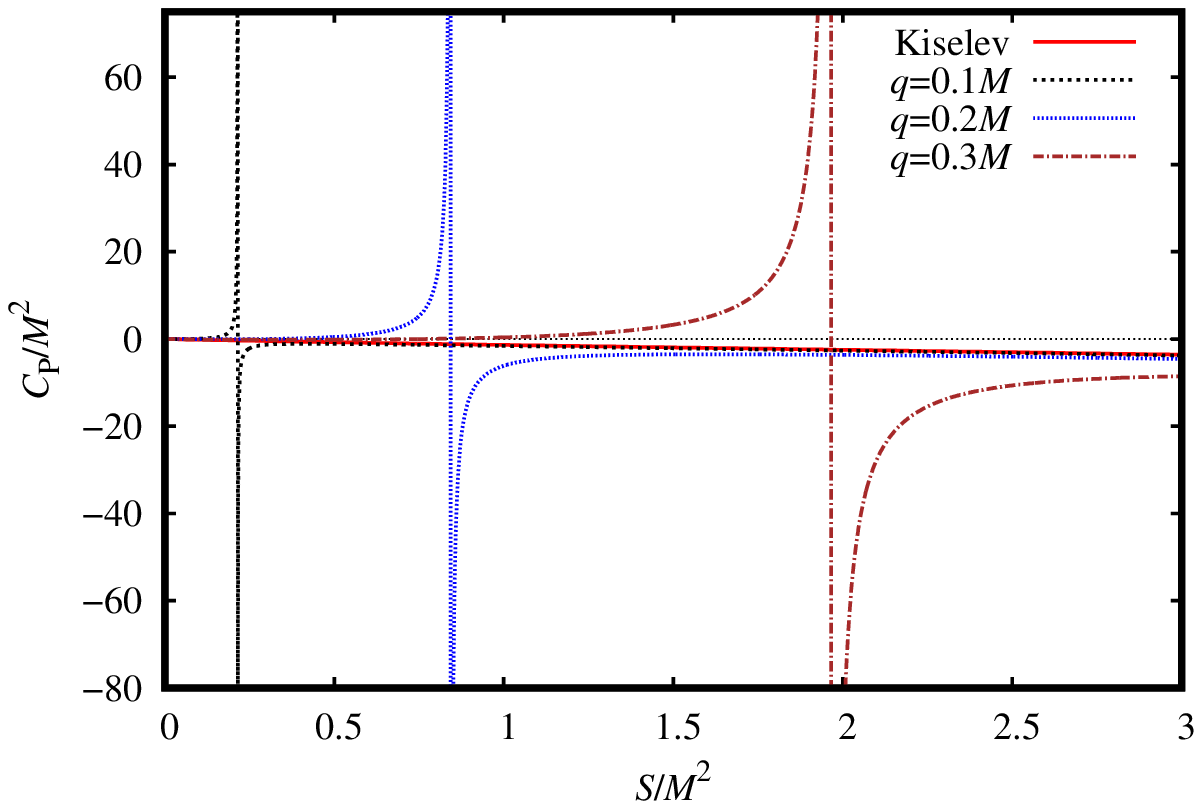}
	\includegraphics[scale=.6]{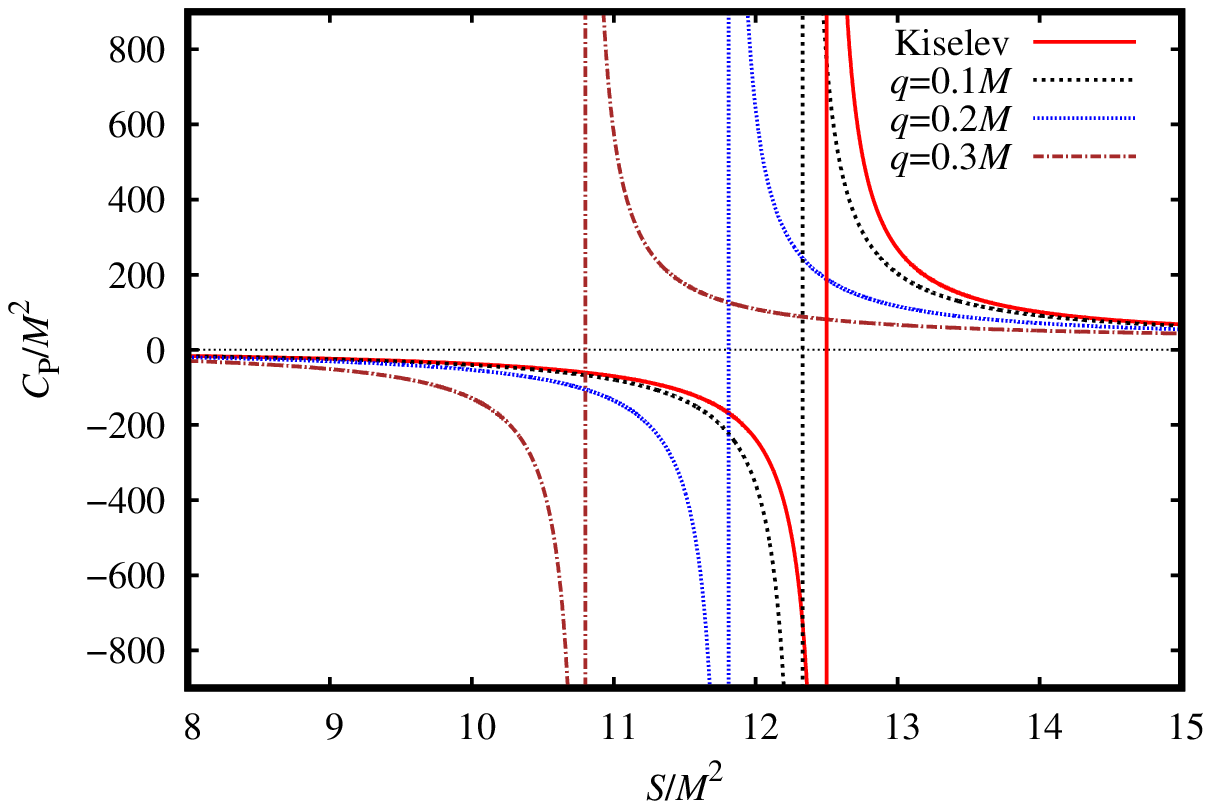}
	\caption{Heat capacity at constant pressure and temperature associated to the solution \eqref{fBK} with $M^2\lambda=-0.08\pi$, $\omega=-2/3$, and $Mc=0.2$.}
	\label{figC2}
\end{figure}

The heat capacity to the Bardeen-(anti-)de Sitter and Kiselev-(anti-)de Sitter solutions are
\begin{eqnarray}
C_{PB}=\frac{2 S \left(\pi q^2+S\right) \left(2 \pi ^2 q^2+\lambda S^2-\pi S\right)}{\pi S^2 \left(3 \lambda q^2+1\right)-7 \pi ^2 q^2 S-2 \pi ^3 q^4+\lambda S^3},\\
C_{ PK }=\frac{2 S \left(-6 c \pi ^{\frac{3 (\omega +1)}{2}} \omega +\lambda S^{\frac{3 (\omega +1)}{2}}-\pi S^{\frac{3 \omega }{2}+\frac{1}{2}}\right)}{6 c \pi ^{\frac{3 (\omega +1)}{2}} \omega (3 \omega +2)+\lambda S^{\frac{3 (\omega +1)}{2}}+\pi S^{\frac{3 \omega }{2}+\frac{1}{2}}}.
\end{eqnarray}
In Fig. \ref{figC1} we see how the heat capacity to the solution \eqref{fBK} behaves. There are both positive and negative values for $C_P$, so the solution is thermodynamically stable for some values of entropy. This behavior is different from the Schwarzschild solution where the heat capacity is always negative, $C_P=-2S$. We also see that in the range of small values of entropy and negative heat capacity, the temperature is also negative and therefore this range has no physical meaning. In the Bardeen-(anti-)de Sitter solution, the first phase transition occurs at a higher temperature value than in the other cases, while the second phase transition occurs at a lower temperature value than in the other cases. In this way, the Bardeen solution has a smaller temperature range for the intermediate phase. From Fig. \ref{figC2} we see that the Kiselev-(anti-)de Sitter solution has only one phase transition, while the solution \eqref{fBK} has two phase transitions. 

\section{Equation of State}\label{sec5}
\begin{figure}
	\includegraphics[scale=.6]{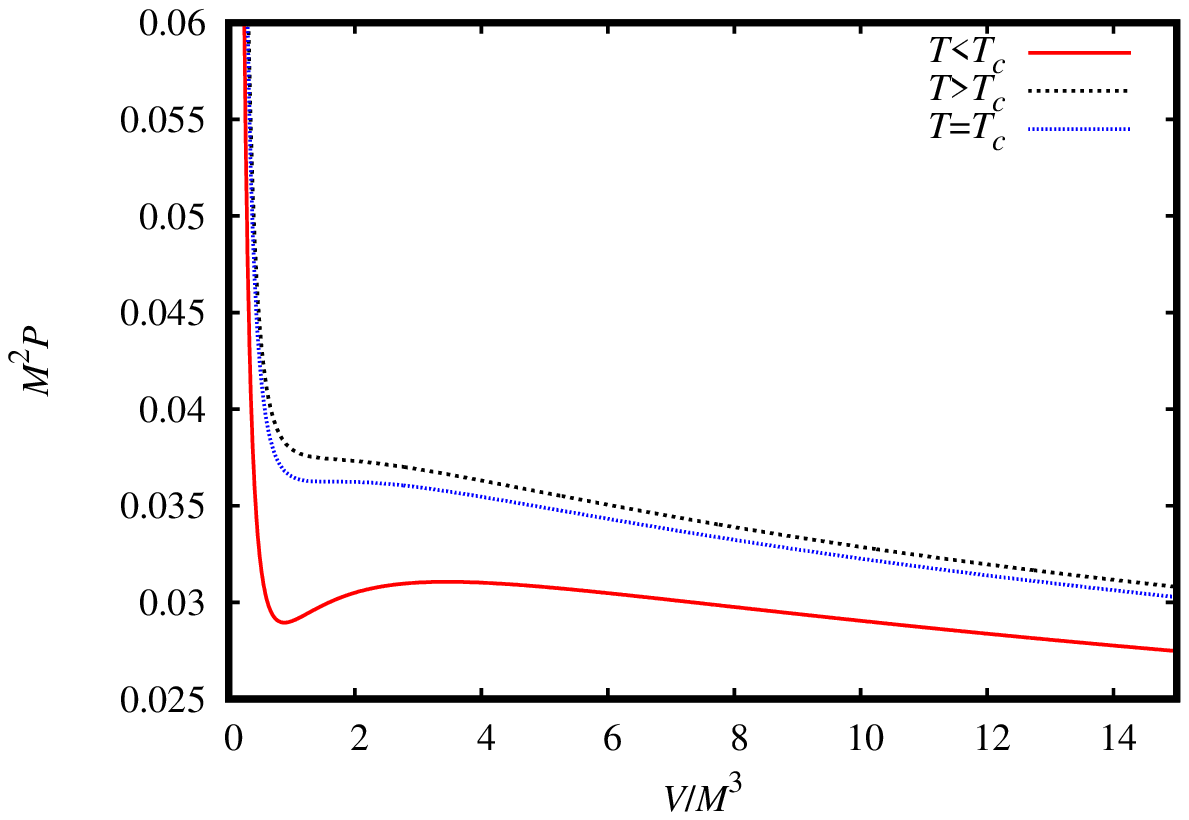}
	\includegraphics[scale=0.6]{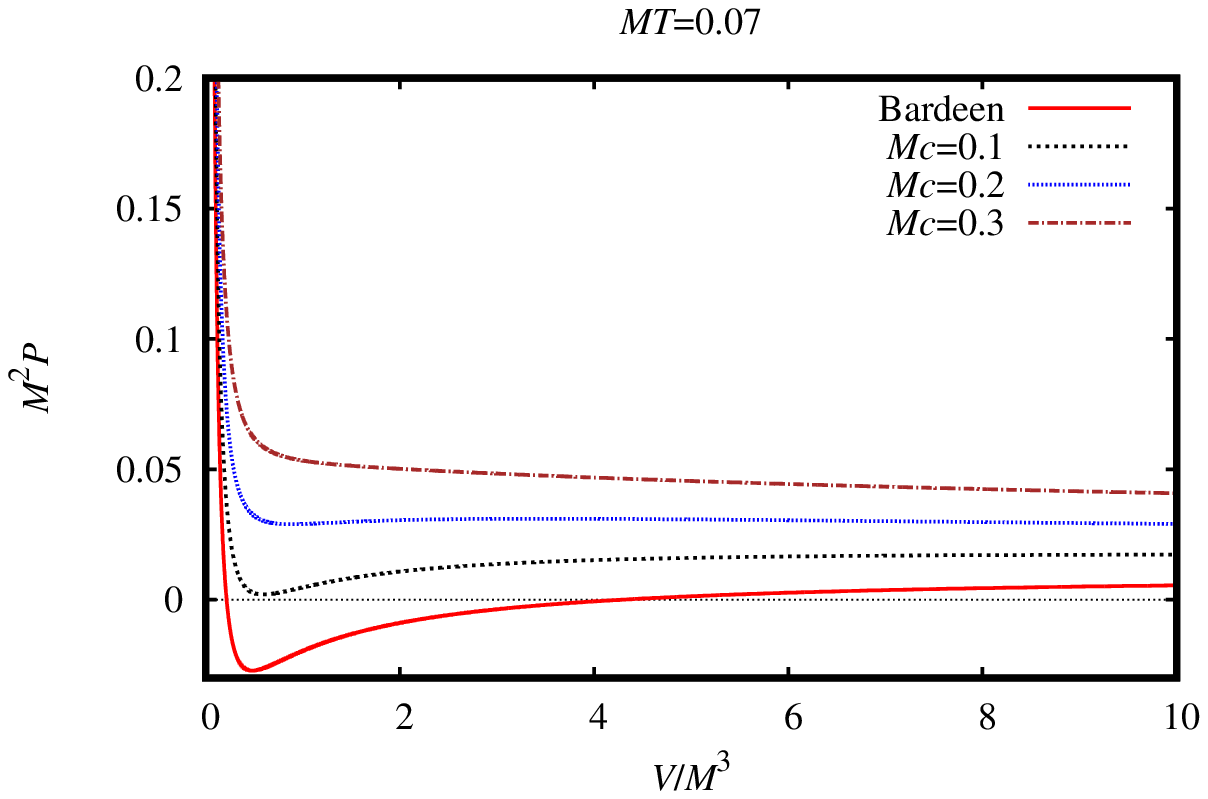}
	\includegraphics[scale=0.6]{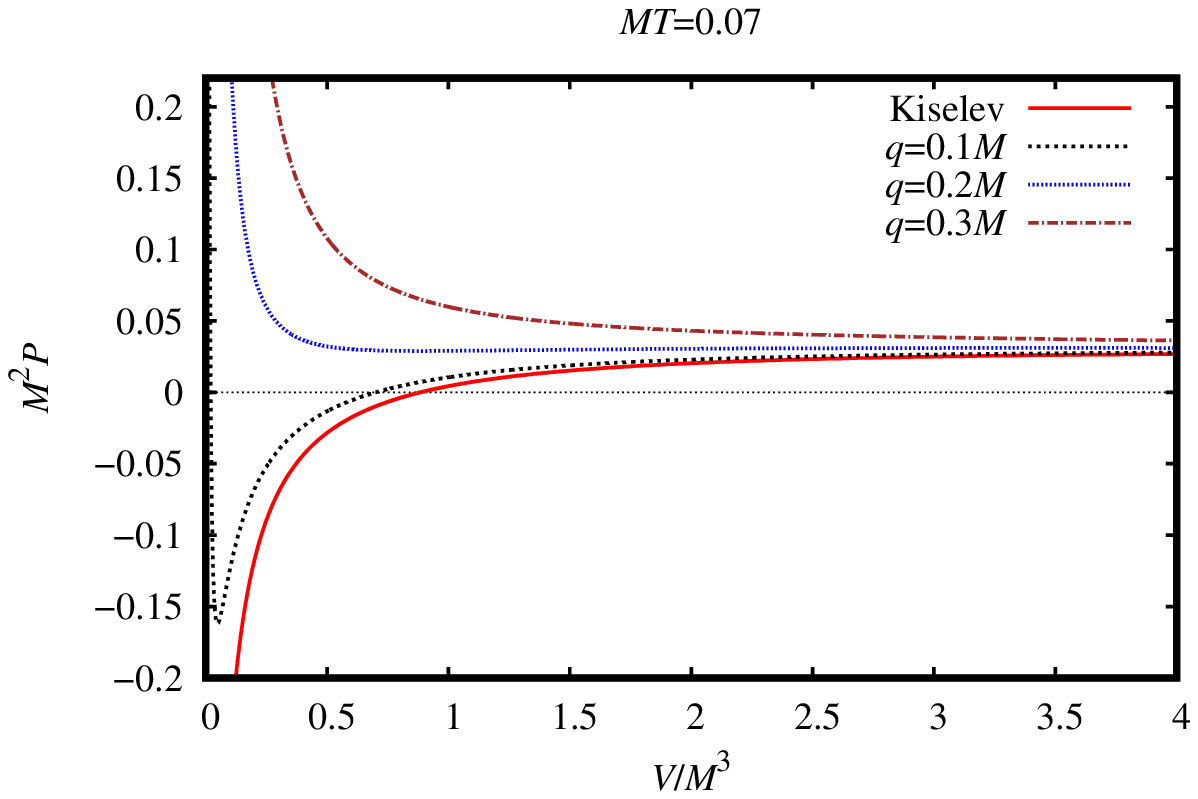}
	\caption{Graphic representation of the $P-V$ behavior of the solution \eqref{fBK} with $q=0.2M$, $Mc=0.2$ and, $\omega=-2/3$. There is also the behavior at a fixed temperature and different values of $q$ and $c$.}
	\label{fig3}
\end{figure}

From the modified first law \eqref{FistLaw}, the thermodynamic volume is given by
\begin{equation}
V=\frac{\partial \mathcal{M}}{\partial P}=\frac{4\pi r^3_+}{3}.\label{Vol}
\end{equation}
With \eqref{Tk} and \eqref{Vol} we obtain the equation of state $P(T,V)$, i.e.
\begin{equation}
P(T,V)=\frac{4 q^2 \left(6 \pi ^{4/3} T \sqrt [3]{V}+(6 \pi )^{2/3}\right)+6 (6 \pi )^{2/3} T V-3 \sqrt [3]{6}V^{2/3}}{36 \sqrt [3]{\pi } V^{4/3}}-\frac{ 4^{\omega } \pi ^{\omega +\frac{2}{3}}c }{3^{\omega+\frac{2}{3}}V^{\omega +\frac{5}{3}}} \left(2 \sqrt [3]{2} q^2 (\omega
+1)+\left(\frac{3V}{\pi }\right)^{2/3} \omega \right).\label{press}
\end{equation}
The pressure to the Bardeen-(anti-)de Sitter and Kiselev-(anti-)de Sitter solutions are
\begin{eqnarray}
P_B(T,V)&=&\frac{\sqrt [3]{\frac{\pi }{6}} \left(2 q^2+3 T V\right)}{3 V^{4/3}}+\frac{2 \pi q^2 T}{3 V}-\frac{1}{2\ 6^{2/3} \sqrt [3]{\pi } V^{2/3}},\\
P_K(T,V)&=&-c \left(\frac{4 \pi }{3}\right)^{\omega } \omega V^{-\omega -1}+\frac{\sqrt [3]{\frac{\pi }{6}} T}{\sqrt [3]{V}}-\frac{1}{2\ 6^{2/3} \sqrt [3]{\pi } V^{2/3}}.
\end{eqnarray}
The critical points result from the conditions \cite{Dehyadegari:2017flm,Dayyani:2017fuz,Wei:2020poh}
\begin{equation}
\left(\frac{\partial P}{\partial V}\right)_T=0, \ \mbox{and} \ \left(\frac{\partial^2P}{\partial V^2}\right)_T=0.\label{Pressurecondition}
\end{equation}
Unfortunately, we are not able to solve this equation analytically, but if we fix some values for parameters $q$, $\omega$, and $c$, we get numerical values for $T_c$, $P_c$, and $V_c$. In Fig. \ref{fig3} we show the isotherm considering three cases, $T<T_c$, $T>T_c$, and $T=T_c$, where $T_c$ is the critical temperature. It is clear that the pressure is higher for higher temperature values. At a fixed temperature, the pressure increases as the charge and $c$ increase. For the Kiselev-(anti-)de Sitter solution it is not possible to solve the condition \eqref{Pressurecondition} because there is no inflection point in $P(V)$. So we are not able to find the critical points.

\begin{figure}
	\includegraphics[scale=0.6]{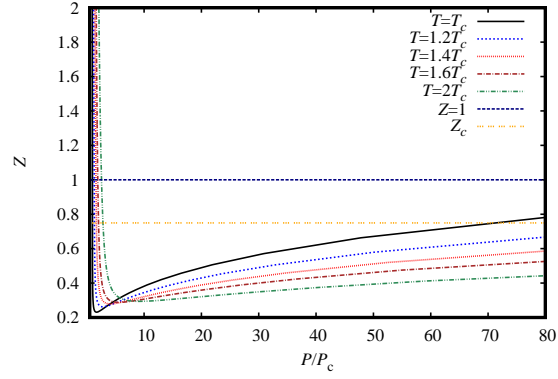}
	\caption{Compressibility factor in terms of the ratio $P/P_c$ to $q=0.2M$, $Mc=0.2$, and $\omega=-2/3$.}
	\label{FigZ}
\end{figure}

Since we have the equation of state $P\left(T,V\right)$, we can obtain the compressibility factor $Z$, which is defined as \cite{Stanley}
\begin{equation}
Z=\frac{PV}{T}.\label{ CF }
\end{equation}
For an ideal gas, i.e., a gas in which there is no interaction between the particles that compose it, we have $Z=1$ \cite{Stanley}. For a real gas where there is interaction between the particles that make it up, $Z$ depends on factors such as pressure and temperature \cite{Stanley}. Thus, the compressibility factor basically tells the deviation of the behavior of a real gas from that of an ideal gas. Since $P\left(T,V\right)$ is a function of temperature and volume, $Z$ is also a function of $T$ and $V$ and is written as
\begin{equation}
Z(T,V)=\frac{4 q^2 \left(6 \pi ^{4/3} T \sqrt [3]{V}+(6 \pi )^{2/3}\right)+6 (6 \pi )^{2/3} T V-3 \sqrt [3]{6}V^{2/3}}{36 \sqrt [3]{\pi V} T}-\frac{ c }{T}\left(\frac{4\pi}{3V}\right)^{\omega +\frac{2}{3}} \left( q^2 (\omega
+1)+\left(\frac{3V}{4\pi }\right)^{2/3} \omega \right).\label{ZTV}
\end{equation}
To Bardeen-(anti-)de Sitter e Kiselev-(anti-)de Sitter we find
\begin{eqnarray}
Z_K(T,V)&=&-\frac{c \left(\frac{4 \pi }{3}\right)^{\omega } \omega V^{-\omega }}{T}-\frac{\sqrt [3]{V}}{2\ 6^{2/3} \sqrt [3]{\pi } T}+\sqrt [3]{\frac{\pi }{6}} V^{2/3},\\
Z_B(T,V)&=&\frac{\sqrt [3]{\frac{\pi }{6}} \left(2 q^2+3 T V\right)}{3 T \sqrt [3]{V}}+\frac{2 \pi q^2}{3}-\frac{\sqrt [3]{V}}{2\ 6^{2/3} \sqrt [3]{\pi } T}.
\end{eqnarray}
In Fig. \ref{FigZ} we see how $Z$ behaves when we increase the pressure for different values of the temperature. The compressibility factor diverges to small values of the pressure. As said, for a real gas $Z\neq 1$, but for small values of pressure some gasses behave like an ideal gas. For a van der Waals fluid, the compressibility factor at the critical point is $Z_c=0.375$ \cite{Dayyani:2017fuz}. For the solution \eqref{fBK} we find $Z_c\approx 0.75$, which is twice as much as in the van der Waals case.

\begin{figure}
	\includegraphics[scale=0.6]{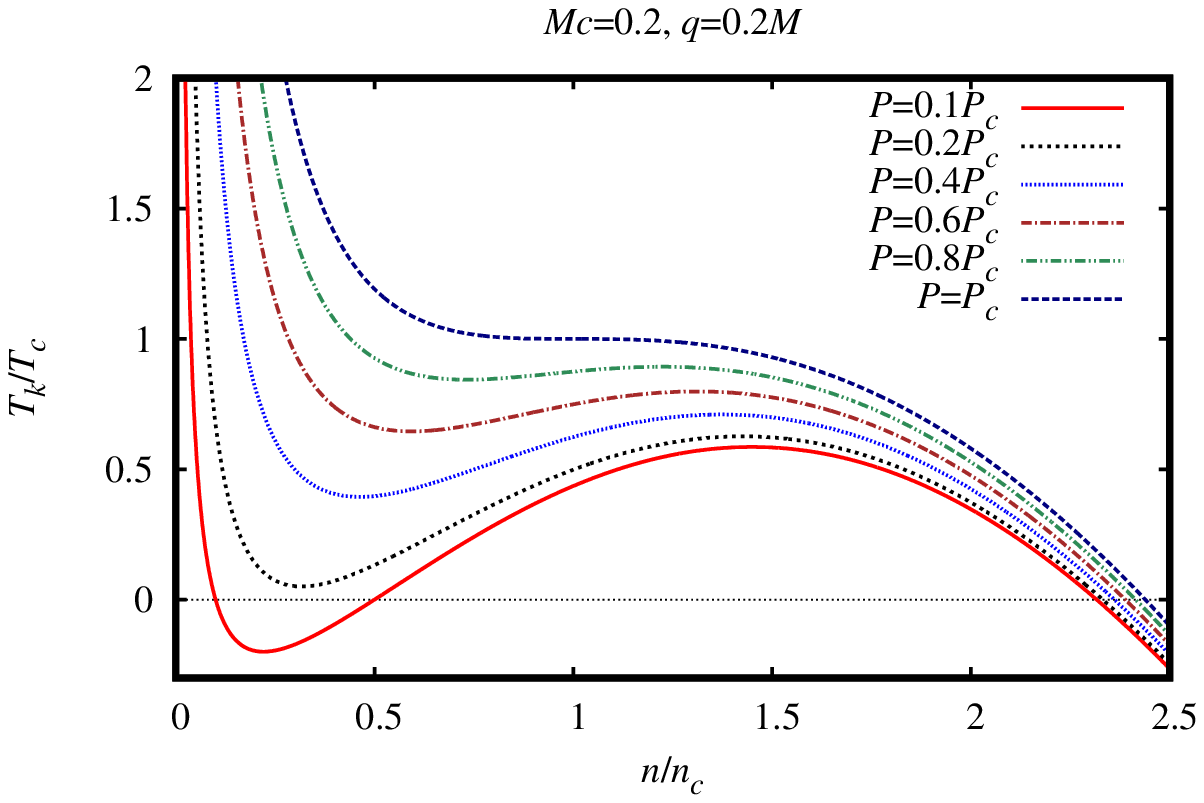}
	\includegraphics[scale=0.6]{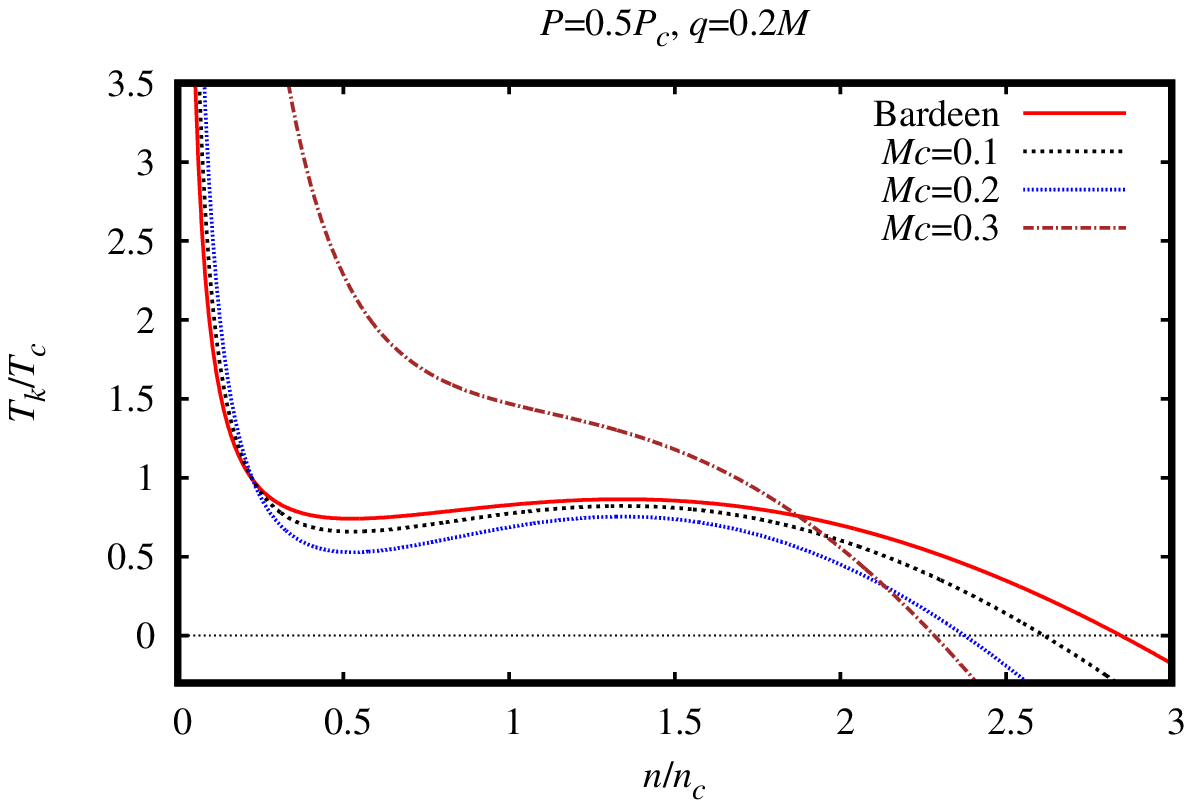}
	\includegraphics[scale=0.6]{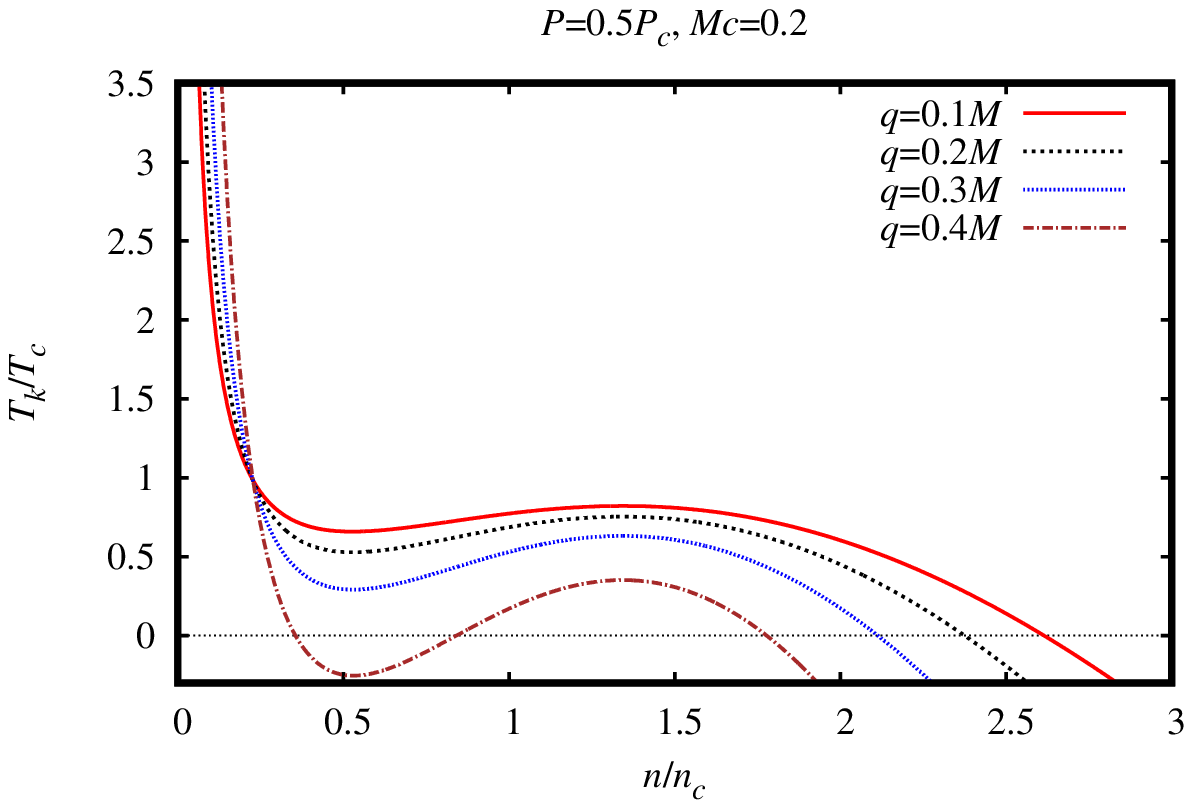}
	\caption{Behavior of ratio $T_k/T_c$ in terns of number density to $\omega=-2/3$.}
	\label{FigTn}
\end{figure}

In the context of fluids, the compressibility factor, when analyzed microscopically, provides information about the interaction of the molecules that make up the fluid \cite{James}. The most important point is this interpretation of the molecules in the context of black holes. According to Wei and Liu, it is possible to analyze a black hole microscopically through the idea of virtual molecules \cite{Wei:2015iwa}. Thus, it is interesting to use the number density of virtual micromolecules of the black hole, i.e.
\begin{equation}
n=\frac{1}{v}=\frac{1}{2 l_p^2r_+},
\end{equation}
where $v$ is the specific volume of the black hole fluid and $l_p=\sqrt{\hbar G/c^3}$ is the Planck length. Since we are considering natural units, $l_p=1$ and $n=1/v=1/2 r_+$. Note that the specific volume is linear with the radius of the event horizon. In terms of the number density, the temperature is
\begin{eqnarray}
T(n)=\frac{3 c\left(2n\right)^{3 \omega +2}n \left(4 n^2 q^2(\omega +1)+\omega \right)-8 n^4 q^2+2 \pi P+n^2}{2 n\pi\left(4 n^2 q^2+1 \right)}.\label{EqTn}
\end{eqnarray}
For Bardeen-(anti-)de Sitter and Kiselev-(anti-)de Sitter we find
\begin{eqnarray}
T_B(n)&=&\frac{-8 n^4 q^2+n^2+2 \pi P}{8 \pi n^3 q^2+2 \pi n},\\
T_K(n)&=&\frac{6 c 2^{3 \omega} \omega n^{3 \omega +2}}{\pi }+\frac{P}{n}+\frac{n}{2 \pi }.
\end{eqnarray}
In Fig. \ref{FigTn} see how the temperature behaves when we change the number density for different values of the pressure. There is a maximum value of $n$ and, depending on the value of $P$, there are other limits in the number density. For example, for small values of pressure, there is a range where the temperature is negative, so these values of $n$ are not allowed. Towards larger values of charge, the number density increases. As $c$ increases, the number density decreases.

\section{Helmholtz free energy and isothermal compressibility}\label{sec6}
To verify the phase transition, we should analyze the Helmholtz free energy \cite{Dehyadegari:2017flm,Dayyani:2017fuz,Wei:2020poh,Liu:2014gvf,Nam:2018tpf,Dehyadegari:2020tau,NaveenaKumara:2020biu}, $F$, given by
\begin{equation}
F=M- TS.
\label{HFE}
\end{equation}
With \eqref{mass}, \eqref{Tk}, and \eqref{Entropy}, we obtain
\begin{equation}
F=\frac{1}{6} \left(\frac{\left(q^2+r_{+}^2\right)^{3/2} \left(\left( 8 \pi P r_{+}^2+3\right) r_{+}^{3 \omega +1}-6 c\right)}{r_{+}^{3 (\omega +1)}}+\frac{3 r_{+}^2 \left(6 c r_{+}^{-3 \omega }-8 \pi P r_{+}^3-3 r_{+}\right)}{2 \left(q^2+r_{+}^2\right)}-\frac{6 c}{r_{+}^{3 \omega }} -3 c (3 \omega +1) r_{+}^{-3 \omega }+3 r_{+}\right).
\label{helm}
\end{equation}
To Bardeen-(anti-)de Sitter and Kiselev-(anti-)de Sitter we find
\begin{eqnarray}
F_B&=& \frac{r_+}{2}-\frac{r_+^3 \left(8 \pi P r_+^2+3\right)}{4 \left(q^2+r_+^2\right)}+\frac{\left(8 \pi P r_+^2+3\right) \left(q^2+r_+^2\right)^{3/2}}{6r_+^2},\\
F_K&=&-\frac{c}{r_+^{3 \omega}}\left(1+\frac{3\omega}{2}\right) -2 \pi P r_+^3+\frac{r_+}{6} \left(8 \pi P r_+^2+3\right)-\frac{r_+}{4}.
\end{eqnarray}

\begin{figure}
	\includegraphics[scale=0.6]{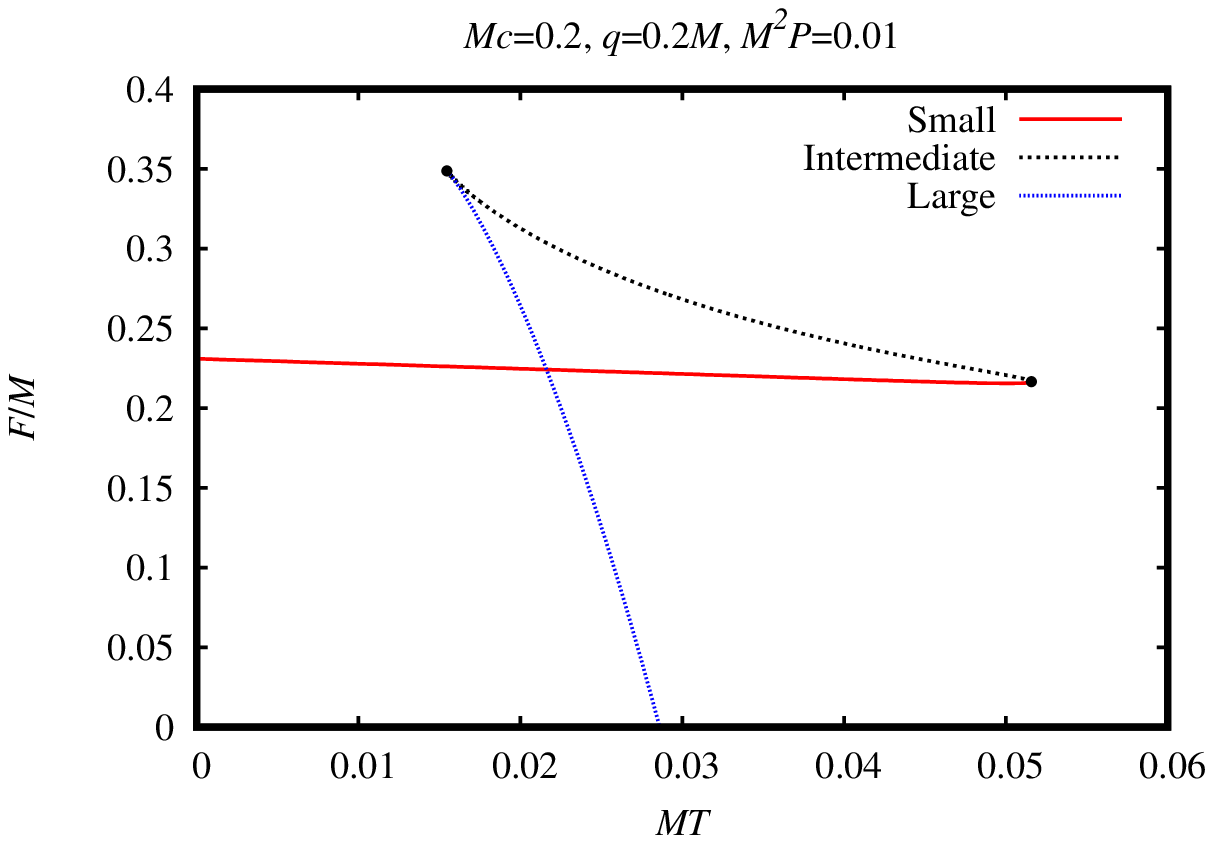}
	\includegraphics[scale=0.6]{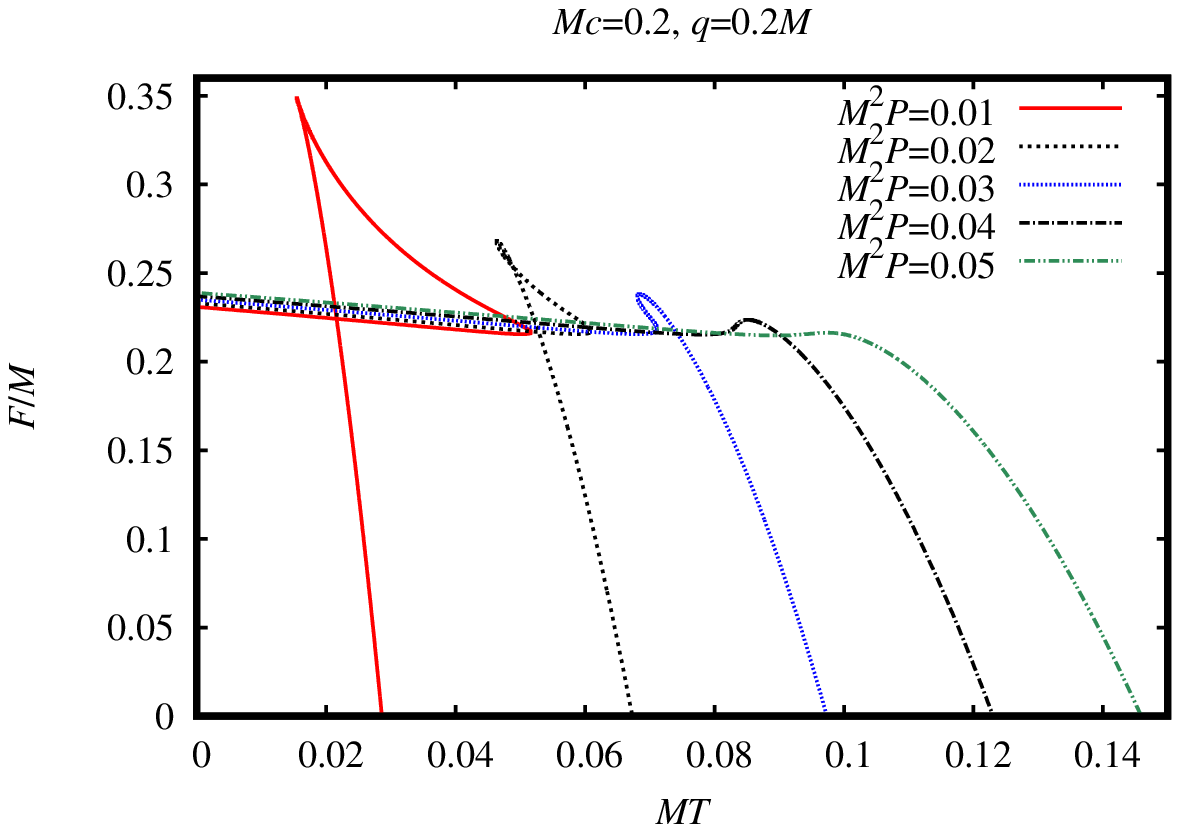}
	\includegraphics[scale=0.6]{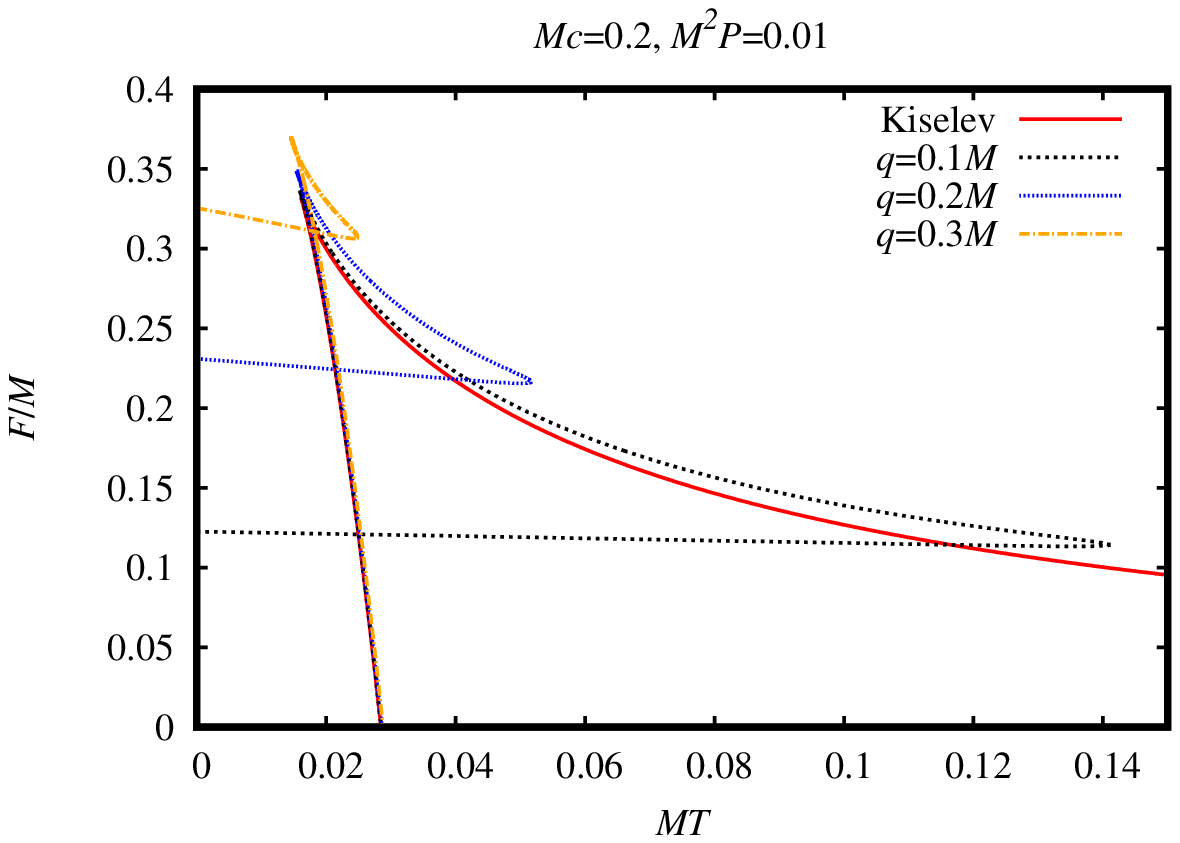}
	\includegraphics[scale=0.6]{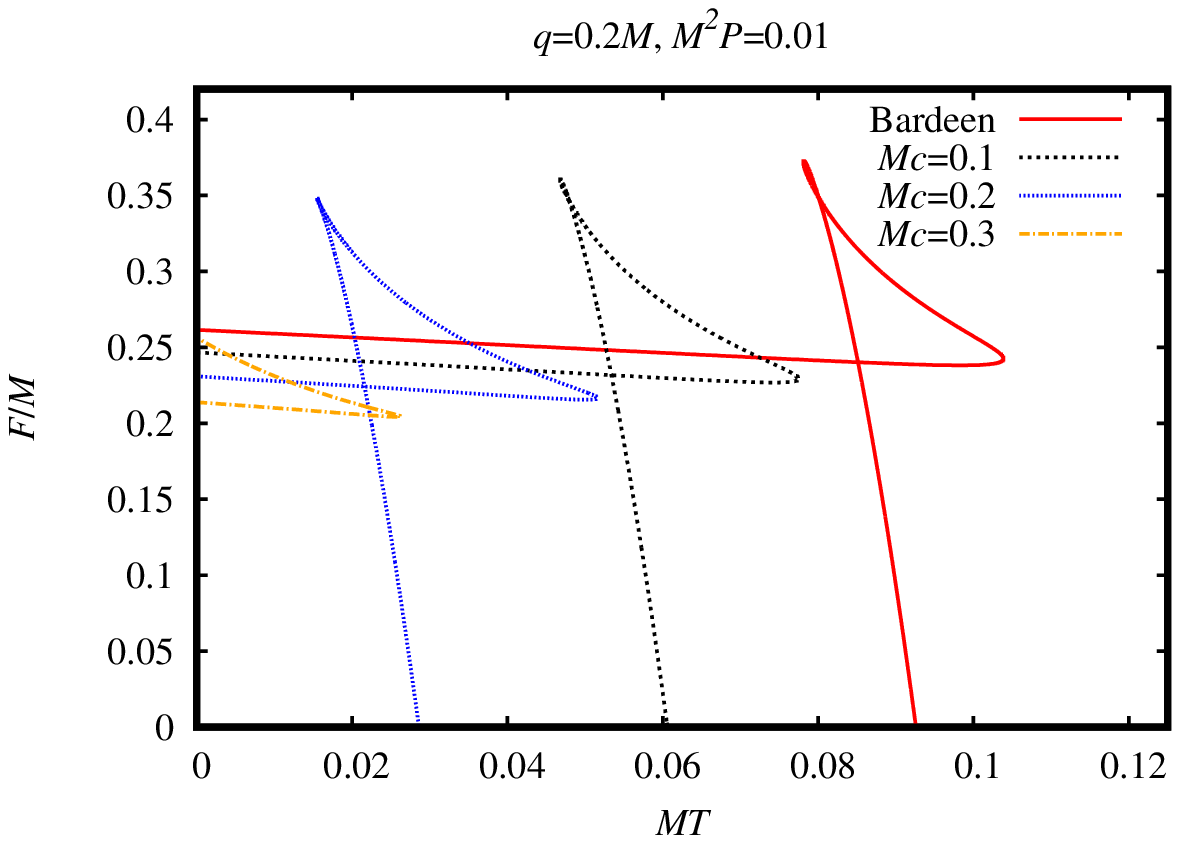}
	\caption{Helmholtz free energy as a function of temperature to $\omega=-2/3$.}
	\label{FigHel}
\end{figure}

In Fig. \ref{FigHel} we show the behavior of the Helmholtz free energy. For small values of the temperature, we have only small black holes. As we increase the temperature, there is more than one possible state. There is a temperature value for which there are three different states, but two of them have the same energy. Interestingly, there is a black hole, the intermediate one, for which the heat capacity is negative, that is, it is thermodynamically unstable. When we increase the pressure, the intermediate black hole disappears and the phase transition occurs from a small black hole to a large black hole. The Kiselev solution has only two phases and as the charge increases, the temperature required for the phase transition decreases. In the Bardeen solution, there are three phases, but as $c$ increases, one of the phases disappears at positive temperatures.

\begin{figure}
	\includegraphics[scale=.6]{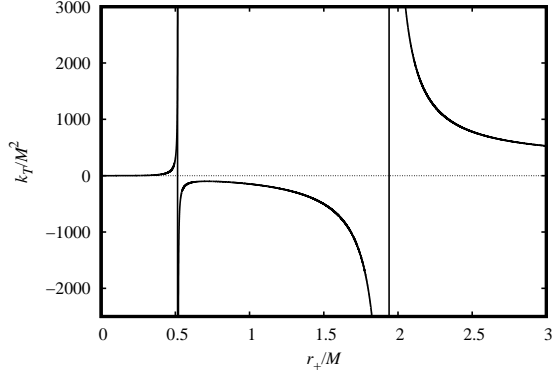}
	\caption{Isothermal compressibility associated to the solution \eqref{fBK} with $q=0.2M$, $M^2P=0.01$, $\omega=-2/3$, and $Mc=0.2$.}
	\label{figkt}
\end{figure}

Another way to analyze the phase transition is through isothermal compressibility. If we use \eqref{compress} with \eqref{press}, we get
\begin{eqnarray}
k_T&=&24 \pi \left(r_{+}^2+q^2\right)r_{+}^{3\omega +5}\left[r_{+}^{3 \omega+1} \left(q^2 r_{+}^2 \left(24 \pi Pr_{+}^2+7 \right)+8 \pi P r_{+}^6+2q^4-r_{+}^4\right)\right.\nonumber\\
&-&\left.6 c\left(q^2 r_{+}^2 (\omega (6 \omega +7)+4)+q^4 (\omega +1) (3\omega +2)+r_{+}^4 \omega (3 \omega+2)\right)\right]^{-1}.\label{kt}
\end{eqnarray}
To Bardeen-(anti-)de Sitter and Kiselev-(anti-)de Sitter we get
\begin{eqnarray}
k_{ TB }&=&\frac{24 \pi r^{4} \left(q^2+r^2\right)}{q^2 r^2 \left(24 \pi P
 r^{2}+7\right)+8 \pi P r^6+2q^4 -r^4},\\
k_{ TK }&=&\frac{24 \pi r^{3\omega +3}}{6 c \omega - (18 c \omega^2 +18 c \omega)+\left(8 \pi P r^2-1\right)r^{3\omega +1}}.
\end{eqnarray}
When the isothermal compressibility is positive, the solution is thermodynamically stable \cite{Stanley}. In Fig. \ref{figkt} we see that the solution is stable for a small and large black. As we already know from the heat capacity, the intermediate black hole is thermodynamically unstable.

\section{Critical exponents}\label{sec7}
As we have seen, some physical quantities such as isothermal compressibility and heat capacity diverge during a phase transition. Even though these quantities diverge, we can analyze their behavior near the transition point. For this purpose, we use the critical exponents \cite{Dayyani:2017fuz,Wei:2020poh}. To obtain the critical exponents, we use
\begin{equation}
t=\frac{T-T_c}{T_c}, \ v=\frac{V-V_c}{V_c}, \ \mbox{and} \ p=\frac{P}{P_c}.\label{pvt}
\end{equation}

The critical exponents $\alpha$, $\beta$, $\gamma$, and $\delta$ are obtained from \cite{Stanley}
\begin{eqnarray}
C_V=T\left(\frac{\partial S}{\partial T}\right) \propto \left|t\right|^{-\alpha},\\
\eta=V_1-V_2 \propto \left|t\right|^{\beta},\\
k_T=-\frac{1}{V}\frac{\partial V}{\partial P}\propto \left|t\right|^{-\gamma},\\
\left|P-P_c\right|\propto \left|V-V_c\right|^{\delta},
\end{eqnarray}
where $\eta$ is the difference in volume between two phases.

Before proceeding, we have analyzed the phase transition graphically for certain parameter values. Thus, in the next analyses, although it is not obvious, we consider the same values as before, $\omega=-2/3$, $q=0.2M$, and $M c=0.2$.

If we use \eqref{pvt} and rewrite \eqref{press} and expand it for small values of $t$ and $v$, we get
\begin{equation}
p=1+t (a-b v)-d v^3+O\left(v^4,tv^2\right),\label{paprox}
\end{equation}
where $a$, $b$, and $d$ are just constants that are combinations of $\omega$, $q$, and $c$. If we consider $t$ as constant and derive $p$ to $v$, we get
\begin{equation}
dp=-\left(bt+3dv^2\right)dv.
\end{equation}
This relation is important to apply Maxwell's area law, which states \cite{Dayyani:2017fuz,Wei:2020poh}
\begin{equation}
\int VdP=0.
\end{equation}
Thus, considering Maxwell's area law and the fact that the pressure is constant at the phase transition, we obtain
\begin{eqnarray}
p&=&1+t (a-b v_1)-d v_1^3=1+t (a-b v_2)-d v_2^3,\\
0&=&\int_{v_1}^{v_2}\left(v+1\right)\left(bt+3dv^2\right)dv.
\end{eqnarray}
Solving these equations yields the nontrivial solution as follows:
\begin{equation}
v_1=-v_2 \propto \sqrt{t} \ \mbox{and} \ \eta\propto\sqrt{t}.
\end{equation}
This means that one of the critical exponents is $\beta=1/2$.

We can also use \eqref{paprox} to calculate the isothermal compressibility and we find that
\begin{equation}
k_T\propto \frac{1}{t},
\end{equation}
such that $\gamma=1$. 

We can rewrite the condition $\left|P-P_c\right|\propto \left|V-V_c\right|^{\delta}$ as $\left|p-1\right|\propto \left|v\right|^{\delta}$. From \eqref{paprox} we get
\begin{equation*}
p-1\propto t (a-b v)-d v^3.
\end{equation*}
However, in the critical isotherm we have $t=0$, so that,
\begin{equation}
p-1\propto v^3.
\end{equation}
With this result, we find the critical exponent $\delta=3$.

To determine $\alpha$, we need to analyze the heat capacity at constant volume. To obtain $C_V$, we need to rewrite the entropy as a function of volume, which is given by
\begin{equation} S(V)=\sqrt [3]{\frac{9V^2\pi }{16}}.
\end{equation}
If the volume is constant, the heat capacity at constant volume is
\begin{equation} C_V=\left.T\left(\frac{\partial S}{\partial T}\right)\right|_V =0.
\end{equation}
This means that the critical exponent is $\alpha=0$.
We see that these exponents satisfy the Griffiths, Rushbrooke and Widom equalities \cite{Wei:2020poh,Stanley,Griffiths};
\begin{eqnarray}
\alpha+\beta\left(1+\delta\right)=2,\ \mbox{Griffiths}\\
\gamma\left(\delta+1\right)=\left(2-\alpha\right)\left(\delta-1\right),\ \mbox{Griffiths}\\
\alpha+2\beta+\gamma=2, \ \mbox{Rushbrooke}\\
\gamma=\beta \left(\delta-1\right),\ \mbox{Widom}
\end{eqnarray}
which says that there are only two independent exponents.

\section{Conclusions}\label{sec8}
In this article, we have obtained and analyzed a magnetically charged solution that mimics the Bardeen regular black hole, magnetically charged, with cosmological constant, and surrounded by quintessence. The Lagrangian of this solution has as a limit in the Bardeen case at $c=0$. We find a metric function that can have up to four horizons depending on the parameters. By analyzing the Kretschmann scalar, we find that the solution is regular: in the whole spacetime up to $\omega=-1$, at infinity of the radial coordinate when $\omega\leq-1$, in $r=0$ up to $\omega\geq-1$. 

It was also possible to analyze the solution from the point of view of black hole thermodynamics. We obtained two different temperatures $T_k$ and $T_H$ reflecting the use of nonlinear electrodynamics, and therefore we modified the first law by the correction factor \eqref{correctionfactor}. Moreover, analyzing \eqref{Masshomogeneous}, we found that the mass is a homogeneous function with a degree of homogeneity $n=1/2$, and thus we found the Smarr formula \eqref{Smarrcorrect}.

We obtain two values for the heat capacity connected by \eqref{CCrelation}. Analyzing Figs. \ref{figC1} and \ref{figC2}, we conclude that $C_P$ can take both positive and negative values as a function of entropy, i.e., there are regions where the solution is thermodynamically stable $(C_P\geq0)$ and in others it is unstable $(C_P < 0)$. The equation of state \eqref{press} behaves like a van der Walls fluid for $q=0.2M$, $Mc=0.2$ and $\omega=-2/3$, and at the critical point the compressibility factor is twice as large as in the van der Walls case. We also construct the temperature as a function of the number density of virtual micromolecules of the black hole, as shown in Fig. \ref{FigTn}. To obtain a positive temperature, $n$ cannot take any arbitrary value, so there must be some bounds on this variable. 

By analyzing the Helmholtz free energy and the isothermal compressibility of the system, we show that the solution has two thermodynamically stable states, namely small and large black holes. During the phase transition, a third intermediate phase occurs, but it is thermodynamically unstable. Finally, fixing the values $q=0.2M$, $Mc=0.2$, and $\omega=-2/3$, we calculated the critical exponents $\beta=1/2$, $\gamma=1$, $\delta =3$, and $\alpha=0$, but only two of them are independent since they are connected by the Griffiths, Widom, and Rushbrooke equalities.

In future work, we hope to explore some topics that are missing in this paper, such as geodesic analysis, shadows, causal structure, maximum analytical extension, and the analysis of the interaction with bosonic and fermionic fields.
 
\vspace{1cm}

{\bf Acknowledgements}: M. E. R.  thanks Conselho Nacional de Desenvolvimento Cient\'ifico e Tecnol\'ogico - CNPq, Brazil, for partial financial support. This study was financed in part by the Coordena\c{c}\~{a}o de Aperfei\c{c}oamento de Pessoal de N\'{i}vel Superior - Brasil (CAPES) - Finance Code 001.


\end{document}